\documentclass[reprint,superscriptaddress,showpacs,amsmath,amssymb,aip,jcp,nofootinbib]{revtex4-1}

\usepackage{dcolumn,bm,book tabs,comment,natbib}
\usepackage[pdftex]{graphicx}
\usepackage[usenames,dvipsnames]{color}
\usepackage{colortbl,longtable}

\definecolor{URLCOL}{rgb}{0,0.52,0.83} 
\definecolor{LINKCOL}{rgb}{0.05,0.5,0} 
\definecolor{CITECOL}{rgb}{0.25,0,0.48} 
\definecolor{PREPRINTCOL}{rgb}{0.0,0.0,0.0} 

\usepackage[pdftex,bookmarks,breaklinks,bookmarksopen,bookmarksnumbered,colorlinks,linkcolor=LINKCOL,linktocpage,citecolor=CITECOL,urlcolor=URLCOL,pdfpagemode=UseOutline]{hyperref}

\definecolor{TITLECOL}{rgb}{0.1,0.2,0.7} 
\definecolor{SECOL}{rgb}{0.1,0.2,0.7} 
\definecolor{CONTENTSCOL}{rgb}{0.1,0.2,0.7} 
\definecolor{SSECOL}{rgb}{0.25,0,0.48} 
\definecolor{SSSECOL}{rgb}{0.2,0.08,0.53} 
\definecolor{FINCOL}{rgb}{0.01,0.3,0.07}

\def\coloredtitle#1{\title{\textcolor{TITLECOL}{#1}}} 
\def\coloredauthor#1{\author{\textcolor{CITECOL}{#1}}} 

\definecolor{URLCOL}{rgb}{0,0.17,0.43} 
\definecolor{LINKCOL}{rgb}{0.05,0.4,0} 
\definecolor{CITECOL}{rgb}{0.35,0,0.48} 

\def\Eqref#1{Eq.~\eqref{#1}}

\def\sec#1{\section{\textcolor{SECOL}{#1}}}
\def\ssec#1{\subsection{\textcolor{SSECOL}{#1}}}
\def\sssec#1{\subsubsection{\textcolor{SSSECOL}{#1}}}
\def\usec#1{\section*{\textcolor{SECOL}{#1}}}

\definecolor{URLCOL}{rgb}{0.1,0.2,0.7} 
\definecolor{LINKCOL}{rgb}{0.1,0.2,0.7} 
\definecolor{CITECOL}{rgb}{0.1,0.2,0.7} 
\def\coloredauthor#1{\author{\textcolor{PREPRINTCOL}{#1}}} 
\def\sec#1{\section{\textcolor{PREPRINTCOL}{#1}}}
\def\ssec#1{\subsection{\textcolor{PREPRINTCOL}{#1}}}
\def\sssec#1{\subsubsection{\textcolor{PREPRINTCOL}{#1}}}
\def\usec#1{\section*{\textcolor{PREPRINTCOL}{#1}}}

\definecolor{lightgray}{gray}{0.8}

\def\W{{\rm ^W}}

\def\F{_{\sss F}}

\def\bea{\begin{eqnarray}}
\def\eea{\end{eqnarray}}
\def\ben{\begin{equation}}

\def\een{\end{equation}}
\def\benu{\begin{enumerate}}
\def\enu{\end{enumerate}}
\def\bei{\begin{itemize}}
\def\eei{\end{itemize}}

\def\br{{\bf r}}

\def\n{n}

\def\sss{\scriptscriptstyle\rm}

\def\x{_{\sss X}}
\def\c{_{\sss C}}
\def\s{_{\sss S}}
\def\xc{_{\sss XC}}

\def\F{_{\sss F}}

\def\LDA{^{\rm LDA}}

\def\GEA{^{\rm GEA}}
\def\GGA{^{\rm GGA}}
\def\acGGA{^{\rm acGGA}}
\def\nGGA{^{\rm RSC}}
\def\PBE{^{\rm PBE}}

\def\TF{^{\rm TF}}
\def\KS{^{\rm KS}}
\def\W{^{\rm W}}
\def\HOMO{^{\rm HOMO}}

\def\unif{^{\rm unif}}

\def\QC{^{\rm QC}}

\def\ac{_{\sss AC}}

\def\half{\frac{1}{2}}

\def\nh{n_{\sss HOMO}}

\usepackage[normalem]{ulem}
\definecolor{darkgreen}{rgb}{0.0,0.66,0.0}
\definecolor{darkred}{rgb}{0.75,0.00,0.0}

\begin{document}
\coloredtitle{
Fitting a round peg into a round hole:
asympotically correcting the generalized gradient approximation for correlation
}
\coloredauthor{Antonio Cancio}
\affiliation{Department of Physics and Astronomy, Ball State University,
Muncie, IN 47306, USA}


\coloredauthor{Guo P. Chen}

\coloredauthor{Brandon T. Krull}

\coloredauthor{Kieron Burke}
\affiliation{Department of Chemistry,
University of California, Irvine, CA 92697, USA}


\date{\today}

\begin{abstract}
We consider the implications of the Lieb-Simon limit for correlation in density functional theory.  In this limit, exemplified by the scaling of neutral atoms to large atomic number, LDA becomes relatively exact, and the leading correction to this limit for correlation has recently been determined for neutral atoms.  We use the leading correction to the LDA and the properties of the real-space cutoff of the exchange-correlation hole to design, based upon PBE correlation, an asymptotically-corrected correlation GGA which becomes more accurate per electron for atoms with increasing atomic number.  When paired with a similar correction for exchange, this acGGA satisfies more exact conditions than PBE.  Combined with the known $r_s$-dependence of the gradient expansion for correlation, this correction accurately reproduces correlation energies of closed shell atoms down to Be.  We test this acGGA for atoms and molecules, finding consistent improvement over PBE, but also showing that optimal global hybrids of
 acGGA do not improve upon PBE0, and are similar to meta-GGA values.
We discuss the relevance of these results to Jacob's ladder of 
non-empirical density functional construction.

\end{abstract}


\maketitle


\sec{Introduction}

A major paradigm
of the development of
density functional theory (DFT) is 
that of the nonempirical application of constraints within
a Jacob's ladder of approximations.
Each rung of Jacob's ladder~\cite{PS01} is characterized by its 
treatment of the exchange-correlation (XC)
energy, the only component of the total energy approximated within the 
Kohn-Sham scheme.
The rungs are to be filled with approximations
that satisfy relevant exact constraints.
An optimal functional at a given rung 
should presumably incorporate the maximum 
amount of information that a functional of that form can.~\cite{SRP15}
Each approximation should improve
over that of lower rungs, usually at higher computational cost.

The ground-level is the Hartree approximation (i.e., XC set to zero);
the first rung is the local density approximation (LDA), whose form
is unambiguously determined by the XC energy of a uniform electron gas.
The local gradient of the density is added at the next rung, 
the generalized gradient approximation (GGA).
For the last two decades, the PBE functional~\cite{PBE96}
has been a popular candidate for this level.~\cite{PGB15}  Its
moderate accuracy for a very broad range of systems is 
because it agrees in large part with the real-space cutoff 
(RSC) construction for a GGA,~\cite{P85,BPW97} and in so doing,
satisfies seven exact constraints.~\cite{PBE96}
The third or meta-GGA rung adds the kinetic
energy density~\cite{Bb98}, 
or alternately, the Laplacian of the density.~\cite{PS94,FT98,CP07,CWW12,MT17}
This rung has been much harder to fill nonempirically,
but recently, the SCAN functional,~\cite{SRP15} constructed with a combination 
of exact conditions and appropriate norms, promises
to become a new standard, overcoming difficulties of 
previous 
attempts.~\cite{PKZB99, TPSS03, PRCC09,  PRCCS11, SHXB13, SXFHHRCSP13}
By the logic of Jacob's ladder, SCAN should
outperform the LDA and non-empirical GGAs like PBE
in almost all areas.  

A problem with the nonempirical approach is that of finding 
effective constraints to optimize a given level of functional.
Finding the optimal constraints to use at a given level is an ill-posed 
problem
-- often the satisfaction of a constraint with a 
lower-rung form requires breaking other, perhaps equally important 
constraints.  
Thus many alternatives to PBE have been 
developed by choosing alternative sets of constraints.~\cite{P91, PRCV08, ZT08a, VMT09, RCS09, CFLD11, CGTV12, CTSCF16,
PCGTV16}
At the meta-GGA level, the flexibility of the form allows many more
constraints to be satisfied, but the problem then is the sheer complexity 
of the form required to do so, and finding enough relevant 
constraints to constrain it.
For the GGA level, because the gradient expansion of the real-space 
hole is known for both X and C, a GGA can be numerically defined by 
cutting off that hole in real-space.  
The exact conditions met by the resulting RSC GGA are largely those that are
implemented in the construction of the PBE.

In this context, the concept of ``appropriate norms" as described in 
Ref.~\onlinecite{SRP15}
takes on importance.  These are paradigmatic systems that a 
density functional at a given level of approximation rigorously satisfies.  
The importance of norms are that they contain more information than other
forms of constraints, and eliminate much of the ambiguity involved in their
application.
The fundamental example is the homogeneous electron gas that 
exactly specifies the LDA.
Unfortunately no such unambiguous norms exist for the GGA level or 
meta-GGA level, although the removal of correlation
self-interaction in single-electron 
systems, a limited norm, is a key target of non-empirical meta-GGAs.

Indeed, the absence of such a norm for a GGA guided the original development 
of constraint-based GGA functionals in terms of the numerical RSC
model for the exchange-correlation hole (which describes XC-induced 
fluctuations in electron density about any electron.)
GGAs capture some general features of this hole, but
lack the capacity to describe the hole of any real system in detail -- 
in a sense similar to fitting a round peg into a square hole.
This limitation underlies the ambiguity in the formulation of 
nonempirical GGAs. 

Over the last decade~\cite{PCSB06, ELCB08, LCPB09, EB09, BCGP16}
 the semiclassical analysis of the electron gas has 
identified what might be considered the most significant norm for DFT. 
An especially fruitful aspect of this approach is the analysis of the limit
in which the external potential and number of electrons are simultaneously
scaled to infinity.~\cite{LS73,LS77,L81,ELCB08}  
This scaling is familiarly manifested by the 
extension of the periodic table of neutral atoms to the limit 
$N=Z \to \infty$. 
Semiclassical
analysis~\cite{S52, S80, S81, ES85, E88, LCPB09, EB09, KR10, BCGP16}
 derives the LDA as the natural limit of this process for any
system and generates an expansion in inverse nuclear charge that
then yields universal corrections to the LDA that
may be satisfied by semilocal GGAs, thus in principle generating
the first two rungs of Jacobs ladder.  
In turn, the fourth-order gradient correction, frequently used in
constructing meta-GGA's, along with higher gradient corrections is 
expected to make a contribution only to higher orders in the
large-$Z$ expansion.~\cite{LCPB09,CTSCF16}
While in simple, one-dimensional 
non-interacting systems such corrections can be explicitly 
derived\cite{RLCE15,RB17},
for real systems, such corrections can at present only be extracted 
numerically, and so far, only for atoms and similar simple cases.

Recent work has provided numerical estimates of these
corrections for the exchange energy and Kohn-Sham kinetic 
energy.~\cite{EB09,LCPB09}
Correlation has awaited
the availability of highly accurate total correlation energies for
a signficant subset of the atoms via quantum chemical methods.~\cite{MT11,MT12} 
Recent work~\cite{BCGP16} has used this data to identify precisely the leading 
energetic correction to
LDA for the correlation energy of neutral atoms.
This correlation constant is likely correct at least 
for non-periodic
Coulombic systems, and perhaps universally,
and we can numerically extract its value for neutral atoms, and hence build
it into approximate functionals.  This is entirely non-empirical, and in
principle, its value could be determined by a long perturbative semiclassical
calculation, as has been done previously at the LDA level for correlation.
A similar (but much simpler) derivation for exchange showed that both the B88
and PBE exchange functionals come quite close to fulfilling the equivalent
exact condition for exchange.~\cite{EB09}

This new information offers a potential resolution to the
issue of finding an appropriate norm for the GGA.
Just as the LDA forms the leading order term in the asymptotic expansion of 
correlation (indeed of any component of the energy)
the GGA is the simplest possible functional which can reproduce
the leading order beyond-LDA term in the expansion, that is, the order
characterized by our recent extrapolations.
Moreover, the process of estimating the high-$Z$ correction to the 
LDA from low-$Z$ data involves constructing a smooth asymptotic form 
that approximates the semiclassical asymptotic expansion for correlation
to all orders of $Z$.  This smooth form, accurately reproducing
quantum chemistry (QC) data for all $Z$,
is in principle exactly fit by a GGA, as we shall show in the course of this 
paper.
Higher rungs of Jacobs ladder appear as
corrections to this smooth form, and generate, for atoms,
rich and complex shell structure effects beyond the scope of this paper.
%
%
We argue then that the high-$Z$ limits of atomic exchange and correlation 
energies and the related approximate smooth asymptotic forms for all $Z$
define an appropriate norm for the construction of the GGA. 
That is, asymptotic analysis produces the ``round hole" that the 
``round peg" of the GGA can (and should) be made to fit.

The purpose of this work then is to construct a GGA-level functional that is 
asymptotically correct -- exact in the large-$Z$ limit 
of neutral atoms.  
A notable parallel in behavior~\cite{BCGP16} between PBE correlation and the smooth
asymptotic trend of QC correlation data
makes PBE the natural reference for constructing an asymptotically correct 
functional.
In the present work, however, we show that in the semiclassical limit there
is a significant contribution to the correlation GGA that is {\em undetermined}
in the PBE derivation, defining a new, eighth constraint, that a
nonempirical GGA should satisfy rigorously. 
By modifying the high density limit of PBE correlation (PBEc), we
enforce this new exact condition on GGA, and agree better with the
high density limit of the real-space cutoff procedure.
This variation on PBE, which we call acGGA
(asymptotically-corrected GGA) has vanishing relative error 
in the non-relativistic limit of large $Z$, and results show that acGGA yields
the most accurate GGA for atomic correlation energies in this limit.
We also develop a corresponding modification to PBE exchange, and find strong
cancellation of errors between X and C for the atoms in acGGA.
The end 
result is a significant improvement over PBE for all atoms
with $Z\!>\!1$.
We test this acGGA for a small set of molecular atomization energies,
showing a moderate and consistent improvement over PBE, showing that
for main-group small-atom molecules, acGGA improves upon
PBE performance.

For real systems, relativistic effects grow
with $Z$ and become indispensable around $Z\!=\!50$ (the precise ground-state
configuation of even Ni depends on them), but this is
beside the point for the present study.  
The available norms, numerical correlation
energies for the homogeneous electron gas and spherical atoms, are specifically
derived
for the nonrelativistic case.
More to the point,
the main lesson of semiclassical analysis is that the $Z\to\infty$
limit has much to say about finite $Z$ atoms, including low
$Z$ where relativistic effects are not important.  

We also note numerous attempts to improve exchange at the
generalized gradient approximation level from constraint-based
considerations~\cite{P91, PBE96, PRCV08, ZT08a, VMT09, CFLD11, CGTV12}
but rather fewer forms~\cite{LM83, P86, P91, PBE96} for
correlation, most notably the early PW91~\cite{P91} and PBE functionals.
This paper answers why this should be the case
in terms of the different asympototic behavior of exchange
and correlation, and particularly the asymptotic behavior
of PBE correlation and its relation
to the real-space cutoff model of the correlation hole.

It is unlikely that acGGA will replace PBE in actual
practice; nevertheless it is vital that each rung of Jacob's ladder 
incorporate the
relevant exact conditions and norms for that rung.  
Here we implement an insight as to what the 
correct GGA rung should look like.  Having
each rung correct is vital for studying the corrections to be included
at the next level. The SCAN functional is unlikely to be
the last word in meta-GGAs, but it includes these asymptotic constraints, 
in a form different from that developed here.~\cite{SRP15} 
We also note a preliminary report concerning asymptotically correcting 
the GGA.~\cite{BCGP14}


This paper is organized as follows.
In Sec.~\ref{Theory} we discuss the theoretical background of our work:
reviewing the asymptotic analysis of the energies of atoms and Lieb-Simon 
scaling, recent findings for correlation, 
the RSC procedure and how it is used to construct PBE.
Sec.~\ref{acGGA} describes the construction of an asymptotically corrected GGA.
In Sec.~\ref{Results}, we test our functional against correlation
energies for the periodic table of atoms and heats of formation of 
molecules, discussing successes (GGA) and limitations (hybrid).
Sec.~\ref{Implications} discusses implications for future density functional 
development, and for understanding the asymptotic limit of atoms, followed by 
conclusions. 

\sec{Theory of asymptotic expansion}
\label{Theory}

  \ssec{The Lieb-Simon limit}
In a landmark 1973 paper, Lieb and Simon proved rigorously that simple
Thomas-Fermi (TF) theory,~\cite{LS73} the precursor to modern Kohn-Sham DFT, 
becomes relatively
exact in a very specific limit, which can be treated with semiclassical
approximations.   In this subsection, we show how that limit can be
approached for any electronic problem, how the various components of
the energy behave in this limit, and how the dominant contributions in
GGA correlation are determined by this limit.

   \sssec{Lieb-Simon scaling}

Lieb-Simon $\zeta$-scaling~\cite{LS77,L81,ELCB08} captures a fundamental 
pattern of the periodic table in a continuous scaling relationship, relating 
this
fundamental intuitive tool of chemistry to a formal mathematical framework.
It is defined as follows: for a system of
$N$ non-relativistic electrons and a one-body potential $v(\br)$,
the $\zeta$-scaled system may be defined as 
                \ben
                  v_\zeta(\br)=\zeta^{4/3}\,
                           v(\zeta^{1/3}\br),~~~~~~~~N_\zeta=\zeta\, N,
                \een
where $1 \leq \zeta < \infty$.
This amounts to scaling the coordinates of the system
while simultaneously increasing the number of particles.  Taking
the potential
    \ben
           v(\br)=-1/r,~~~~v_\zeta(\br) = -\zeta /r,
   \een
and setting $N\!=\!1$ corresponds to mapping the Hamiltonian
of a neutral hydrogen atom to a neutral atom of nuclear
charge $Z\!=\!\zeta$.  Note that in this case, $\zeta$ is 
a continuous generalization of $Z$.

Crucially for our work, Lieb and Simon show~\cite{LS73,LS77} that the
Thomas-Fermi energy is the rigorous limit of the electronic energy
-- as $\zeta \to \infty$,
     \ben
        \lim_{\zeta\to\infty} \frac{E(\zeta)-E\TF(\zeta)}{E(\zeta)} \to 0.
     \een
This holds for nuclear potentials and more generally a large class 
of external potentials that have bound states.~\cite{FLS15} 

Thus $\zeta$-scaling extracts the simplest
possible density functional theory, Thomas-Fermi theory, from
any starting point, however complex.~\cite{DerivNote}
Note that this process does not produce a simple coordinate scaling of the
ground-state charge density.  For example, transforming one atom
into another necessarily generates differences in shell structure.
However, as $\zeta \to \infty$, this shell structure becomes
vanishingly small and the density $\n_\zeta$ of the scaled
system tends to the
Thomas-Fermi limit:      
\ben
\n_\zeta(\br) \to \n\TF_\zeta(\br)
=\zeta^2 \n\TF(\zeta^{1/3}\br).
   \label{eq:ntf}
      \een
Here $\n\TF(x)$ is a smooth, universal scaling form, 
normalized to one.  It does not have a simple closed form, but has 
been recently accurately
parametrized for atomic potentials in Ref.~\onlinecite{LCPB09}.

The importance of this scaling limit for DFT is not hard to discover:
it is universal, applicable to any starting potential, and thus has
universal consequences for DFT.  Moreover, it 
rigorously probes perhaps the most important benchmark for 
DFT development, the periodic table.

   \sssec{Application to neutral atoms}

Lieb and Simon's $\zeta$-scaling takes on quantitative significance with the
technique of asymptotic expansions of the energy of $\zeta$-scaled
systems and in particular, of atoms versus $Z^{-1}$ in the
large-$Z$ limit.~\cite{S52, S80, S81, ES85, E88, LCPB09, EB09, KR10, BCGP16}
Such expansions present the possibility of
a direct systematic derivation of DFT approximations, as
an expansion in a small parameter.~\cite{BCGP16}
And, although proven rigorously for the difficult case of Coulomb-interacting
systems, the results are straightforward to generalize to other, smoother 
potentials.~\cite{FLS15}

Since all systems weakly tend to the TF energy and density in the
Lieb-Simon ($\zeta\!\to\!\infty$) limit, 
TF theory necessarily determines the leading order term in $\zeta^{-1}$, 
in the asymptotic expansion for the energy.
Corrections to the TF energy and density in the universal density
functional must then appear in subsequent orders in the expansion.
Fortunately, it is often the case that the higher the power of $\zeta$ in
the asymptotic series, the simpler the functional form that
can contribute to it. 
This gives one a way to model corrections
such as the gradient expansion (GE) in isolation, albeit
with some complications for the Coulomb potential.~\cite{SemiNote}
Ultimately, at $\zeta\!=\!1$, the full complexity of DFT is revealed.
Thus, we expect that contributions to each order in the expansion 
in $\zeta$
can be captured by successively higher rungs in a mathematically
derived Jacob's ladder of non-empirical approximations.

The power of this approach is revealed by its accuracy.  Applied to 
neutral atoms, where $\zeta$ is equal to the nuclear charge $Z$, and
taking only the leading order Thomas-Fermi term in the
asymptotic expansion of the total energy,  
one predicts the total energy of Rn within 3\%, 
He to within 12\%, and H to within 50\% (and much better
if spin-polarization is allowed for).
The expansion behaves exactly as a perturbation expansion should -- 
the leading order, though clearly not good enough for thermochemistry,
gets the ballpark answer
for \textit{any} $Z\!>\!1$, that is, the entire periodic
table, and including even the next higher-order term makes the
expansion much more accurate.
Thus, the $Z\to\infty$ limit provides the foundation
of the description of matter for any $Z$.

Over the years,\cite{T27,F28,S52,S80,S81,ES85,E88,LCPB09,EB09,KR10,BCGP16}
the asymptotic expansion of the
various contributions, $T\s$, $E\x$, $E\c$ to the total energy in KS theory
have been worked out for the case of atoms.
In the limit $Z\to\infty$ we have
    \bea
        T\s(Z) &=& A\s\, Z^{7/3} - Z^2/2 + B\s Z + \dots, \nonumber\\
        E\x(Z) &=&  -A\x\, Z^{5/3} + B\x Z + \dots, \nonumber\\
        E\c(Z) &=& -A\c\, Z \ln Z + B\c Z + \dots.
    \eea
Here $A\s \approx 0.768745$ as originally derived by Thomas and Fermi,~\cite{T27,F28}
$A\x\approx 0.220874$,~\cite{LCPB09}
and $A\c\approx 0.02073$.~\cite{KR10,BCGP16}
(We use atomic units (energies in hartrees) and give derivations for
spin-unpolarized systems for simplicity.)
As with the total energy,
each leading order term is exactly given by the corresponding local density
approximation in the high density limit, applied to the Thomas-Fermi
density.
For the kinetic energy, this is simply the Thomas-Fermi energy, constructed from
an energy density that behaves as $ n^{5/3}(\br)$, for exchange, the LDA form
$\sim n^{4/3}(\br)$ and for correlation, the high-density limit of LDA
correlation.~\cite{KR10}
Thus the LDA is the large-$Z$ limit for the single atomic
potential and it is plausibly the universal
large-$Z$ limit for electronic matter.

For correlation, the high density limit of LDA was derived by
Gell-Mann and Brueckner~\cite{GB57} who
applied the random phase approximation (RPA) to find:
         \ben
             \lim_{r_s \to 0} \epsilon\unif\c = \gamma \ln{r_s} + \eta,
             \label{epsunif}
         \een
where $r_s\!=\!(3/(4\pi\n))^{1/3}$ is the Wigner-Seitz radius
of density $\n$, $\gamma\!=\!0.031091$. Within the RPA, $\eta\!=\!0.07082$ 
and is $0.04664$ in the exact high-density limit.
(We use an accurate modern parametrization that contains
these limits,~\cite{PW92} here and in construction of GGAs.)
Then~\cite{KS65}
    \ben
           E\c\LDA[\n] = \int d^3r\, \n(\br)\ \epsilon\c\unif(\n(\br)),
           \label{EcLDA}
    \een
which overestimates the magnitude of the correlation energy
of atoms by a factor of two or more.
Now apply Lieb-Simon scaling to this result, by inserting
  $n\TF_Z(\br)$ [Eq.~(\ref{eq:ntf})] and the high density limit
for $\epsilon\unif\c$ [Eq.~(\ref{epsunif})] into Eq. (\ref{EcLDA}), to find:
        \ben
              E\c\LDA = - A\c Z\, \ln Z\, + B\c\LDA Z + ...,
         \een
The leading term can thus be determined as $A\c=2 \gamma/3=0.02073$.
The next term requires a numerical calculation over the 
TF unit density for atoms [Eq.~(\ref{eq:ntf})], yielding
$B\c\LDA =-0.00451$.

     Second-order terms require beyond-LDA density
     functional corrections whose strength depends on
     the properties of the potential being scaled.
     $B\x$~\cite{EB09} is entirely determined by the gradient expansion approximaton
     (GEA) for slowly-varying densities, given by total energy:
          \ben
                 E\x = E\x\LDA + \Delta E\x\GEA,
          \een
     and energy per particle 
          \ben
                 \Delta\epsilon\GEA\x = \mu s^2 \epsilon\LDA\x,
          \een
     with $s = |\nabla n|/4 k_F n$ a measure of inhomogeneity for exchange
     relative to the fermi wavevector $k_F=(3\pi^2n)^{1/3}$.
     The validity of this form for atoms is justified by the fact that $s^2$
     scales as $Z^{-2/3}$ under Lieb-Simon scaling so that higher-order
     gradient corrections such as $s^4$ vanish relative to it.
     However, the value for $\mu$ is different for potentials with and
     without a Coulomb singularity -- that~\cite{KL88} of a
     sinusoidal potential (10/81) is roughly half that which is obtained by
     extrapolating
     the exchange energy of atoms to the large $Z$ limit.~\cite{EB09,CTSCF16} 
     This discrepancy explains the frequent rejection in 
     modern GGAs (both
     empirical and non-empirical) of the
     formally derived parameter of 10/81 for values that approach that of 
     the large-$Z$ limit.~\cite{EB09}

   \sssec{Correlation: Determining $B\c$}
     The second-order term for correlation, $B\c$ is much harder to
     determine than $B\x$ because it is nearly the same order of magnitude as
     the leading correlation term and thus hard to extract from atomic data.
     Moreover, as discussed in the next section where we delineate the careful
     construction of a high-density GGA, the gradient expansion (GE) alone does not suffice to
     describe this coefficient.  At a minimum a GGA is required.
     Nevertheless recent work has determined an accurate estimate
     of $B\c$,~\cite{BCGP16} based on 
     coupled-cluster calculations for closed-shell atoms up through
     $Z=86$,~\cite{MT11} and all atoms up through $Z=54$.~\cite{MT12}
     These, along with the earlier benchmark set~\cite{CGDP93}
     have made possible a reasonably accurate extrapolation
     of $B\c$.

It will be important to describe the extrapolation method in detail
as it generates a benchmark that we use to produce an asymptotically
correct GGA.
As $A\c$ is exact for atoms,~\cite{KR10} we reformulate the
asymptotic expansion to define the target for any beyond-LDA DFT:
        \ben
           B\c = \lim_{Z\to\infty} e\ac(Z),~~~
          \label{Bdef}
          e\ac(Z) = \frac{E\c(Z)}{Z}+A \ln Z,
         \een
     or alternatively as
        \ben
           \Delta B\c = B\c - B\c\LDA =
               \lim_{Z\to\infty} \frac{[E\c(Z) - E\c\LDA(Z)]}{Z}.
          \label{Bdefalt}
         \een
A natural procedure to eliminate the effects of shell structure is to consider
the trend down a specific column of closed shell atoms like the noble gases.
One may find an even smoother trend by averaging over closed 
shells across a single row, as described in Ref~\onlinecite{BCGP16}.
The results are conveniently parametrized versus the inverse of the row
number (which we take to be the principle quantum number of the highest
occupied energy shell, $\nh$.)

\begin{figure}[htb]
\includegraphics[width=0.9\columnwidth]{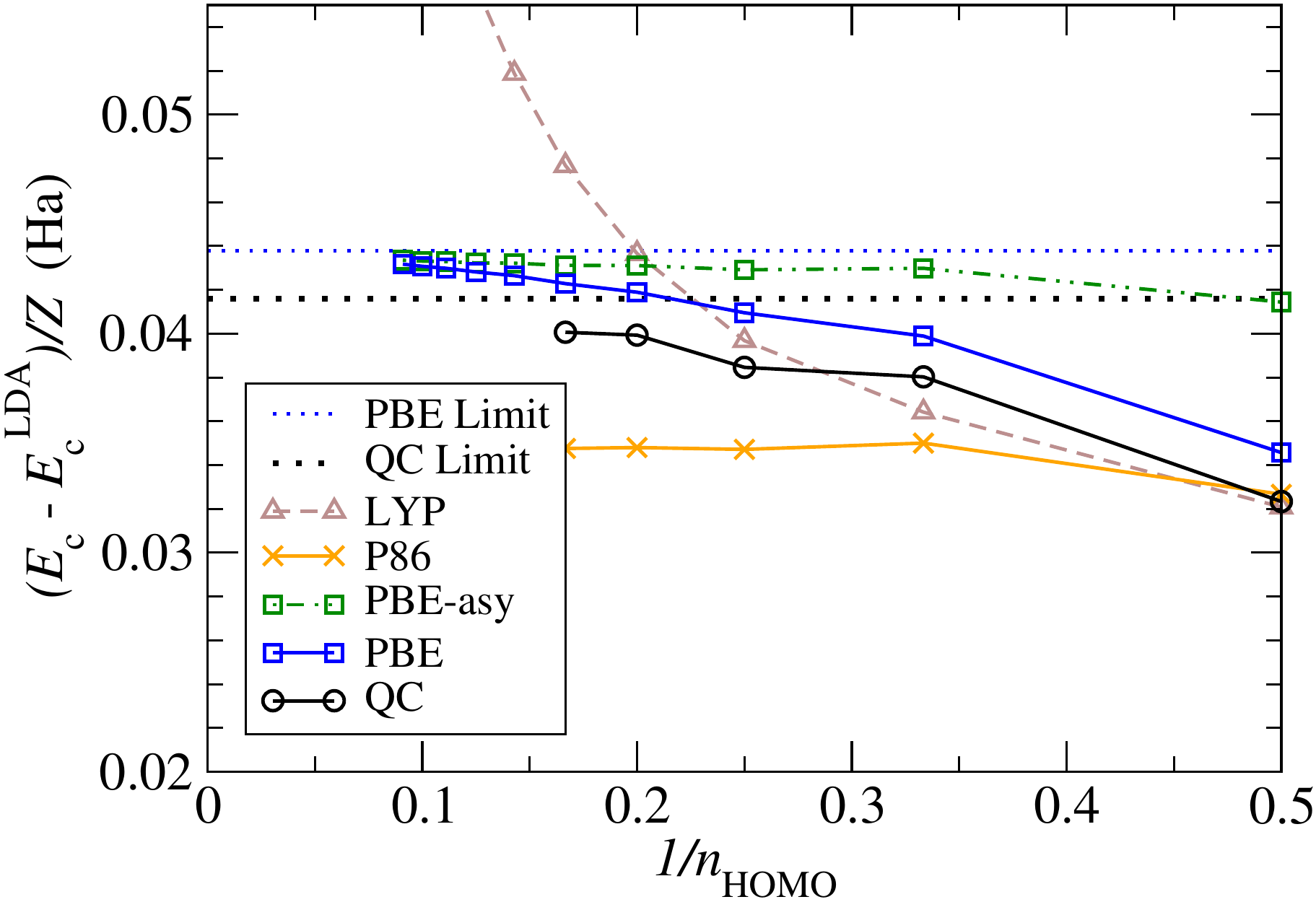}
\caption{
Beyond LDA contribution to the correlation
energy per electron for several GGA approximations compared to accurate
quantum chemistry calculations (QC), averaged over the alkali earth
and noble gas atoms of each row of the periodic table and plotted
versus inverse row number
TF shows the asymptotic limit of PBE,
$B\PBE\c$, 
and PBE-asy is the $r_s=0$ limit of PBE evaluated with a
self-consistent Kohn-Sham density.
}
\label{asyGGA}
\vskip -0.3 cm
\end{figure}

The result of this procedure is shown in Fig.~\ref{asyGGA}, and is compared
to the predictions of several GGA functionals.  The GGA functionals are
calculated out to $\nh=11$, ignoring issues of nuclear stability, in
order to verify their convergence properties in the Lieb-Simon limit.
We see that PBE trends quickly to a $Z\to\infty$ value of
$\Delta B\c = 43.87$~mHa determined by applying the TF density
to the beyond-LDA component of PBE.
In comparision, $B\c$ for the LYP~\cite{LYP88} clearly diverges,
and that of
P86,~\cite{P86} while finite, falls off from the QC trend.
PBE closely parallels the QC data, and assuming
that electronic structure effects grow smaller for larger $Z$,
this parallel is hypothesized to continue on
to the $Z\to \infty$, $1/\nh \to 0$ limit.
The difference may be fit to a straight line trend,
     \ben
          \frac{(E\c - E\c\PBE)}{Z} = -0.00220(38) + \frac{0.0002(13)}{\nh}.
          \label{eqdelBc}
     \een
Thus $\Delta B\c\QC = 41.7$~mHa, shown as the second
horizontal line in Fig.~\ref{asyGGA}.~\cite{BcNote}
This formula becomes a smooth function of $Z^{1/3}$ as $Z\to \infty$
(the difference in $Z$ between an alkali earth or noble of the same row disappears
relative to $Z$ in this limit) and being a constant,
should be largely independent of specifics of the parametrization method.
It clearly reproduces trends in the QC data beyond our initial
target, $B\c$.  In fact, we make a reasonable guess at the smooth contribution
of all the higher-order terms in the asymptotic series.  

The goal of the current paper is the natural followup of this result --
to understand why PBE correlation works as well as it does, and then make it
asymptotically correct.  It allows us to make a precise
(though not exact) definition of
asymptotically correct at the level of a GGA.  The asymptotically corrected
functional should recover the correct value of $B\c$, and as far as possible,
do so by repeating the smooth asymptotic trend to low $Z$ extracted from
QC data.  Thus we have a benchmark that a GGA can be expected
to match -- the smooth asymptotic trend with $Z$ of the periodic table,
eschewing the full details of atomic electronic structure, or the
complexities of covalent bonding of molecules.

\ssec{How (and why) PBE correlation works}

In
order to understand why PBE should be accurate in the Lieb-Simon limit,
we review the history of non-empirical GGAs.  A major role is played by 
the real-space cutoff (RSC) model of the exchange-correlation hole  which
functions as the equivalent of a norm used to generate the PBE and impose
the constraints which it satisfies.  
We also note ambiguities in the high-density limit of RSC that will guide our 
correction to the PBE.

\sssec{The gradient expansion for correlation}

A first step in developing a nonempirical GGA for correlation is the derivation
within the RPA by Ma and Brueckner (MB)
of the leading gradient correction for the correlation energy of
a slowly-varying electron gas.~\cite{MB68}  Define
\ben
   \Delta E\c = E\c - E\c\LDA = \int d^3r\, \n(\br) H\c\left[
r_s(\br),t(\br)\right],
   \label{eqdelEc}
\een
where $t= |\nabla \n|/(2 k_s \n)$ is a dimensionless measure of inhomogeneity 
appropriate for correlation, 
and $k\s=2 (3\n/\pi)^{1/6}$ is the TF screening wavenumber.~\cite{PBE96}
The MB gradient expansion approximation yields
\ben
   H\c\GEA(t) = \beta\, t^2,~~~~~(r_s\to 0)
   \label{eqMB}
\een
with $\beta=0.066725$.  This so strongly overcorrects $E\c\LDA$ for
atoms~\cite{MB68} that we find that $E\c$ becomes positive
for all atoms.
MB showed that a simple Pad\'e
approximant works much better, creating the first correlation GGA, and
inspiring the work of Langreth and Perdew,~\cite{LP80} among others.

We can apply Lieb-Simon scaling to the gradient expansion to
show \textit{a priori} the unsuitability of the GE for the density functional
description of correlation, and thus the need for a GGA.
As $Z \to \infty$, the GE applied to the TF density scales as $Z\ln{Z}$,
not $Z$, giving a spurious gradient correction to $A\c$ in the
asymptotic expansion, as shown in Appendix A.
Only a GGA gives a gradient correction that scales correctly.
The divergent behavior in the LYP estimate of $B\c$
seen in Fig.~\ref{asyGGA} is in part caused by the use of the
simple gradient expansion form.
At small $Z$, the GE corrections are tempered by deviations from the
homogeneous electron gas form of the LDA to produce an excellent
description of correlation, but the cost is a necessary failure at
large $Z$.  

\sssec{Real-space hole construction of the GGA}
Underlying the PBE and related GGAs is the non-empirical real-space
 cutoff (RSC) model for the XC hole\cite{P85,PBW96,BPW97c}, so we review
it in detail.  
It will serve as the foundation for asymptotically correcting PBE. 

The XC hole is defined as
\ben
\n\xc(\br,\br')=\int_0^1d\lambda\, (P_\lambda(\br,\br')/\n(\br)-\n(\br'))
\een
where $P_\lambda(\br,\br')$ is the pair probability density at coupling
constant $\lambda$ along the adiabatic connection curve.  Then
      \ben
             E\xc = \half\int d^3r\, \int d^3r'\,
                 \frac{\n(\br)\, \n\xc(\br,\br')}{|\br-\br'|}.
            \label{Exchole}
      \een
It is interpreted as a change in density at $\br'$ given
an electron observed at $\br$, and may be constructed by taking
the adiabatic integral over coupling constant for this quantity.~\cite{HJ74, LP75, GL76} 
$E\xc$ does not depend sensitively on the details
of the XC hole, but rather on its system and angle average:
     \ben
         E\xc = \half\int 4\pi u^2du\, \frac{1}{u} \langle\n\xc(u)\rangle
     \een
with $u = |\br-\br'|$, and the average is over the other coordinates in
Eq.~\ref{Exchole}.
The XC hole obeys important normalization sum rule
that $\int d^3r'\, \n\xc(\br,\br') = -1$
while the correlation hole alone obeys $\int d^3r'\, \n\c(\br,\br') = 0.$

The LDA can be considered as
approximating the true XC hole by that of a uniform gas:
\ben
   \n\xc\LDA (\br,\br') = \n(\br)\, \left[\bar g\unif\left(r\s(\br),|\br-\br'|
             \right) 
          -1 \right]
    \label{nxcLDA}
\een
where $\bar g\unif$ is the pair-correlation function
of the uniform gas.~\cite{PW92b}
Insertion of this approximate hole into Eq. (\ref{Exchole})
yields $E\xc\LDA[\n]$.
While $\epsilon\xc\unif(\n(\br))$ is not accurate
point-wise,\cite{JG89} (that is, it is not comparable to the integral
over $r'$ in Eq.~\ref{Exchole}) the
system and angle average of the LDA hole is.
This is because the LDA hole satisfies basic conditions -- it
obeys the particle sum-rules for both exchange and correlation
and satisfies the negativity condition for exchange, $n\x(\br,\br') \leq 0$.
So it mimics the exact hole.  Conversely, the exact $E\xc$ depends
only upon the system average of the exact hole, and this is insensitive to
details of electronic structure of an inhomogeneous system, so capturing
these major features suffices.
Hence the reliability and systematic errors of LDA.~\cite{JG89}

XC hole analysis also shows why the gradient expansion fails:
$\n\xc\GEA$ for a sufficiently rapidly varying system has large unphysical
corrections to $\n\xc\LDA$,
violating the exact conditions that the LDA obeys.\cite{BPE98}
For correlation, the correction to the LDA hole~\cite{PBW96}
is proportional to $t^2$ and positive definite, a
response to the averaged exchange hole, which becomes deeper and
more localized, and therefore more efficient at screening.
The GEA hole thus must break the normalization sum rule for correlation
for any non-zero $t$, and do so drastically for situations in which $t^2$
diverges. 

By restoring these exact conditions, the RSC construction determines
the very difficult and essentially nonlocal
piece of information needed to reproduce the hole of an atom or molecule
-- its finite range.
For correlation, the GEA hole is made to satisfy the zero sum-rule by
cutting it off outside a finite radius $v_c$.
This crude procedure is surprisingly effective at predicting
the finite range of real holes, while the GEA hole is typically very good
at small interparticle distances, dramatically improving upon the 
LDA hole in this limit.~\cite{CF12}

\sssec{Constraints derived from the real-space cutoff}
The RSC model, through Eq.~\ref{Exchole},
defines a numerical GGA that naturally generates the properties and constraints
that PBE and related functionals attempt to meet, and suggests
a robust functional form.  In this sense, it can be considered the generator
of PBE and motivates our using it to generate its correction.
The constraints can be separated into the high density limit $r_s=0$,
of immediate interest to us, and those that impose corrections
for finite $r_s$.

In the $r_s=0$ limit, and at low $t$,
the RSC by construction reduces to the Ma-Brueckner GE form, Eq.~(\ref{eqMB}).
At high $t$, typical of a finite system, the RSC cut-off procedure removes the
logarithmically divergent LDA energy term in Eq.~(\ref{epsunif}), $\gamma \ln{r_s}$,
yielding a finite $E\c$.
This is the limit reached by the uniform scaling of the density of any
finite system to high density, here the correlation energy is constrained to 
be bounded from below.~\cite{L89,L91}
RSC also gives a physically reasonable interpolation between the two for
finite $t$.

At finite density, the low-$t$ limit also reduces to the Ma-Brueckner
gradient expansion,
ignoring the weak dependence of the coefficient $\beta$ on $r_s$
calculated by Langreth and coworkers~\cite{HL85, LP80} and
Rasolt and Geldart.~\cite{RG86, GR76}
The high-$t$ limit yields a diverging
GEA correlation hole with a sum rule so unphysically positive that the
RSC procedure cuts it off almost entirely.  This
yields an $E\c$ that vanishes as $1/t^2$,~\cite{BPW97c} satisfied by
requiring $H_c(r_s, t\!\to\!\infty) \!\to\! -\epsilon\c\LDA$.
At very low density, such as the asymptotic tail of a finite
system, this limit is reached for almost any value of $t$.  For atoms,
it may be helpful to think of this as 
the finite-$t$, low density ($r_s \!\to\!\infty$) limiting case,
complementary to the $r_s\!=\!0$ limit discussed above.
The PBE correlation functional, like its predecessor PW91,~\cite{P91,BPW97c}
is based on a simple analytic parametrization of this numerical GGA, and
attempts to capture not only its limit cases but the entire range of
dependence on $r_s$ and $\zeta$.

In the high density limit, the correction to the
LDA reduces to a function of the single variable $t$, and both PBE
and PW91 use the
simplest possible form that can satisfy both high- and low-$t$ limits:
    \ben
         H\c(0,t) = \gamma \ln{(1 + T^2)},
         \label{PBEasy}
    \een
defined in terms of a rescaled inhomogeneity parameter:
    \ben
         T = \sqrt{\frac{\beta}{\gamma}}\, t.
        \label{eqT}
     \een
PW91 adds a second piece to the RSC correlation, in order to ensure a 
zero exchange-correlation correction in the linear response limit.  
Unfortunately, this term reduces to the gradient expansion in the $r_s\to 0$
limit, and like the gradient expansion diverges unphysically.  We drop this 
second piece in our discussion.

\begin{figure}[htb]
\includegraphics[width=0.9\columnwidth]{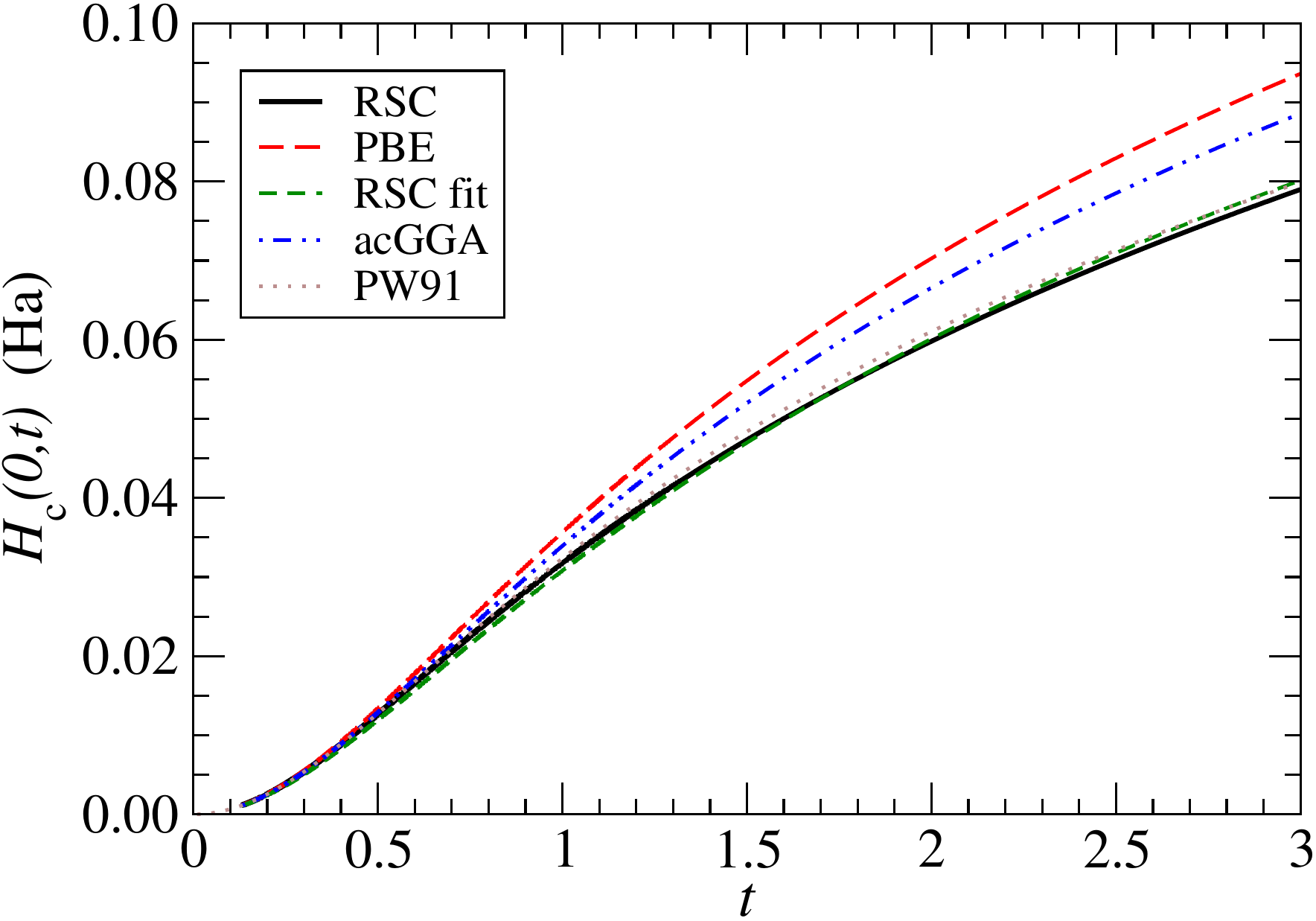}
\caption{
The asymptotic ($r_s\!=\!0$) of the beyond LDA component $H\c(0,t)$ for
generalized gradient approximations, including the
the real-space cutoff (RSC), 
the RSC contribution to PW91, PBE, our fit to the
RSC using Eq.~(\ref{PBEnewasy}), and the asymptotically corrected GGA (acGGA).
}
\label{figRSC}
\vskip -0.3 cm
\end{figure}

Fig.~\ref{figRSC} shows $H\c$ as a function of $t$ for PBE and
the RSC contribution to PW91,
in comparison to the numerical RSC.
The PW91 adjusts $\gamma$ from the RPA value
by a modest amount so as to give a close match to the numerical RSC at
finite $t$.
In doing so it sacrifices the constraint of a finite correlation
energy at high $t$.
In contrast, PBE preserves Ma-Brueckner low-$t$ correlation and
the RPA value for $\gamma$.
But greater attention to the limiting values of $t$ within this restricted
form creates a modest mismatch with the RSC at finite $t$.  The
PBE is thereby justifiable purely on constraints in \textit{limiting}
cases, obviating ultimately the need for reference to the XC hole.

At finite $r_s$, the analytic parametrization of the RSC generalizes to
           \ben
                H\c\PBE(r_s,t)=\gamma\ln{\left ( 1 + T^2 f\c(y)\right )}.
           \een
This defines a cutoff function $f\c(y)$ with a form
\ben
    f\c(y) = (1 + y)/(1 + y + y^2),
    \label{fcpbe}
\een
where
   \ben
      y = a(r_s) T^2
      \label{ypbe}
   \een
identifies the transition between
high and low density behaviors, and the form of $f(y)$
approximates the behavior of the numerical RSC.
It determines $a(r_s)$ implicitly by enforcing
zero net correlation energy in the large $y$ limit:
            \ben
           H\c(r_s\!\to\!\infty,t) = \gamma\ln(1 + T^2/y)
                                      = -\epsilon\c\LDA(r_s)
             \label{pbe}
            \een
which is satisfied by
          \ben
             a(r_s) = \{\exp\left[-\epsilon\c\LDA(r_s)/\gamma\right] - 1
                      \}^{-1}
                  \label{eqars}
          \een
The function $a(r_s)$ is roughly linear in
$r_s$, so that $y \sim s^2$, the scale invariant exchange
inhomogeneity parameter.  

Before moving on, we consider the message of 
the RSC asymptotic form Eq.~(\ref{PBEasy}). 
Early correlation functionals, such
as LYP but also Perdew 86~\cite{P86} and Langreth-Mehl~\cite{LM83}, experimented with a wide variety of forms in this limit, but lacked knowledge of the proper limit of correlation in the limit of uniform scaling
to high density, with $r_s\to 0$ and $t\to \infty$ simultaneously.
The result was either a divergence in this limit for LYP, or a rather poor estimate for $B\c$, 31.0~mHa for Perdew 86.
The part of PW91 that is based on the RSC improves upon the
LDA in this limit, improving $B\c$ to 34.6.  
PBE drops the divergent part of PW91 and fully implemented the uniform limit, yielding a nearly correct $B\c$ of 39.38.  What we need ($\sim 37.2$) is a modest
improvement upon what is already provided in Eq.~(\ref{PBEasy}) by enforcing the correct
constraint under uniform scaling.

\sec{Constructing a new high density GGA}
\label{acGGA}
In this section we show how the PBE construction fails to fully determine
the leading correction to LDA in the Lieb-Simon limit for correlation.
This correction can be folded in to PBE correlation, while still respecting
all conditions PBE correlation was designed to satisfy.  We also show how
including density-dependence in this limit is largely irrelevant, so we do
not do so in our acGGA.  Finally, we discuss which exchange GGA should be
coupled with acGGA.

    \ssec{The high-Z limit of the GGA}

We begin by exploring
the implication of taking the combination
of the high-density limit of the RSC model \textit{and} high-$Z$
together.  To do so, we define an asymptotic PBE by taking the
high density limit of PBE for all densities,
         \ben
             \epsilon\c^{aPBE}(r_s,t) = -\gamma \ln{r_s} + \eta +
                    H\c^{aPBE}(r_s,t)
         \een
where the first two terms are the high density limit of the LDA
[Eq.~\ref{epsunif})], and the GGA correction is

         \ben
             H\c^{aPBE}(r_s,t) = \gamma
                  \ln{\left(1 + T^2 \right) },
            \label{PBEasyalt}
         \een
applied for all $r_s$.
We include calculations using this form in Fig.~\ref{asyGGA}.

The coefficient $B\c\PBE = 39.36$~mHa for PBE is simply the
expectation of this high-density form of PBE using the TF density,
and is shown as a dotted line.  This term \textit{alone} predicts the full
self-consistent PBE model within $80\%$ down to $Z=4$ ($\nh = 2$),
showing the power of asymptotic analysis.
Evaluated with self-consistent densities, aPBE (green squares)
gives the correction to $B\c$ due to the change in density from
TF case.  This is seen to be a small effect for all $\nh$.
The difference of aPBE and PBE shows the effect of the
finite-density correction to $H\c$, and naturally turns on mostly for
the first two values of $n\HOMO$.  But crucially, it is
almost perfectly zero for larger $Z$, where the small change
due to the change in density is dominant.  The low-density
correction of PBE is only relevant for the lowest
rows of the periodic table and QC data mimics this behavior closely.
Changing $B\c$ alone, i.e., modifying aPBE
to retrieve the QC value of $B\c$ (black dotted line) promises
therefore to reproduce the QC values for most of the periodic table.

\begin{figure}[htb]
\includegraphics[width=0.85\columnwidth]{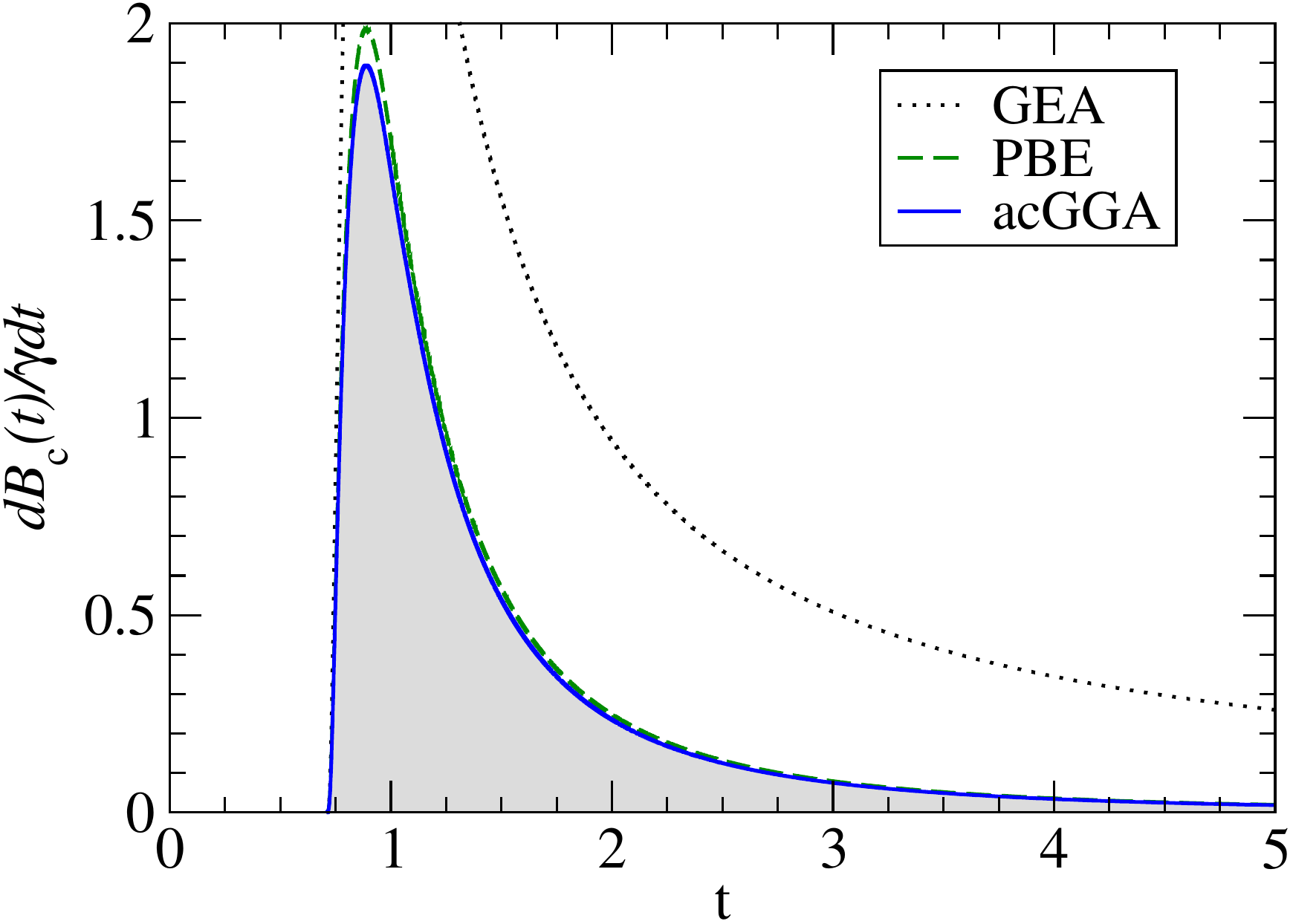}
\caption{Plot of $B\c$ represented as an integral over the
inhomogeneity parameter $t$.  Obtained parametrically by plotting
$4\pi r^2 n(r) H\c\left(0,t(r)\right) / (\gamma dt/dr)$ versus $t(r)$
evaluated for the TF density.
Green dashed line represents the asymptotic limit of PBE; black dotted,
the GEA; blue, the acGGA.  Shaded area is the integral $B\c\acGGA - B\c\LDA$.
}

\vskip -0.3cm
\label{figBct}
\end{figure}

Secondly, we illuminate the nature of the asymptotic
constraint we wish to use and how it affects the form of the GGA.
Take the asymptotic expansion coefficient $\Delta B\c$ as expressed by Eq.~(\ref{Bdefalt}), and the value for it, 0.0417 Ha, extrapolated from QC data.  Insisting that this condition be met by a GGA with gradient correction given by the general form of Eq.~(\ref{eqdelEc}) leads to the following constraint
      \ben
          \frac{1}{Z}
               \int d^3r\, n\TF_Z(\br) H\c\left(0,t[n\TF_Z(\br)]\right)
                     = B\c\!-\!B\c\LDA \sim 0.0417
          \label{eqBc}
          \een
where $B\c\LDA =-0.00451$ and $n\TF_Z$ is given by Eq.~(\ref{eq:ntf}).
We examine the values of $t$ that contribute to this integral,
by changing the integration variable in Eq.~(\ref{eqBc}) to $t$ to
obtain the function $dB/dt$, shown in Fig.~\ref{figBct}.
The curve as shown thus integrates to $B\c$.
The GEA is clearly too large in magnitude and has a slowly decaying $1/t$ tail
that leads to a logarithmic divergence.
The RSC asymptotic form implemented in PBE removes most of the correction
of the GEA and particularly the high-$t$ tail.  It thus obtains a distribution
strongly peaked around $t\sim 0.9$, no values of $t$ smaller than 0.72,
and a rapidly decreasing tail for $t>1$, with a greater than $95\%$ contribution
to $B\c$ for $t<5$.  Thus $B\c$ basically pins down the value of
$H\c(0,t)$ for the characteristic Thomas-Fermi value of $t\sim 1$.
The needed asymptotic correction is a small $(\sim 5\%)$ reduction of
this curve in order to reduce $B\c$; and the solution we describe below,
the acGGA, is shown here as well.

In contrast, asymptotic scaling for exchange tells us about $s \!\to\! 0$: 
$s \!\sim\! \sqrt{r_s} t$ and goes to zero for finite $t$ as $r_s \!\to\! 0$.
Asymptotic analysis of this limit conflicts with the
conventional small-$s$ expansion about the uniform gas,~\cite{KL88}
indicating why disagreement with \textit{a priori} calculations proved
desirable for density functional description of real systems.~\cite{PCSB06,EB09}
Asymptotic scaling for correlation tells us about $t\sim O(1)$, a genuinely
new piece of information in addition that of the limit of uniform scaling
to high density, and the gradient expansion.  Thus it does not
necessarily conflict with prior results, indicating why keeping the
Ma-Brueckner gradient expansion for correlation was not a problem
for the development of realistic density functionals.

   \ssec{Correcting the high-Z limit of PBE}
As discussed above, 
we expect that correcting the leading order term $B\c$ in the
asymptotic expansion for correlation will play a dominant role in
reducing the PBE correlation error for all $Z$.
On the other hand, the accuracy of $B\c\PBE$ suggests
the real-space cut-off procedure from which it derives is highly
accurate at high density.
We thus construct an asymptotically correct GGA by extending the analytic
RSC form to give flexibility to match low-$t$, high-$t$ and $t=1$ behaviors
independently.
We do this by modifying Eq.~(\ref{PBEasy}) to 
    \ben
        H\c\acGGA(0,t) = \gamma \ln{(1 + P(t) T^2)},
         \label{PBEnewasy}
    \een
where 
    \ben
             P(t) = (1 + t/\tau)/(1+ \tilde c t/\tau).
             \label{eqPt}
    \een
To determine a suitable choice of
parameters for $P$ we first fix both $\tau$ and $\tilde c$ to match
the numerical RSC without the limitations of Eq.~(\ref{PBEasy}).
Keeping both the large-$t$ coefficient $\gamma$ and small-$t$
coefficient $\beta$ at the RSC values, we match the
second order term in the RSC large-$t$ limit -- the finite constant
that is left after cancelling the spurious $\ln{r_s}$ divergence
in the LDA correlation.  As derived in Appendix B, this condition
is satisfied by $\tilde c = 2.4683$.  We then set
$\tau=4.5$ to match the RSC curve at finite $t$, and show the
result, labeled ``RSC fit" in Fig.~\ref{figRSC}.
This model yields a value of $B\c\nGGA$ of 0.0327, somewhat off from our 
extrapolated value,
and reflects the uncertainty in RSC in this limit.

To construct an approximation without this uncertainty, we
keep $\tau$ the same, but choose $\tilde c\ac \!=\! 1.467$, which reproduces
our best estimate of $B\c = 0.0372$.~\cite{CacNote}
This result, an asymptotically correct GGA, lies between the RSC
and PBE GGAs, as shown in Fig.~\ref{figRSC}.
This indicates the good quality
of the original RSC for finite $t$, but also indicates that the
PBE was a step in the right direction.

   We make $P(t)$ a function of $t$ not $t^2$ in order to
   match the high-$t$ limit.  This alters the low-$t$ gradient
   expansion, producing a new term proportional to $t^3$.
   The practical effect of this is very small for the asymptotically
   correct model for $P(t)$
   ($\tilde c\ac$) as the third-order coefficient is nearly zero.

    \ssec{Extension to finite density}
        To construct an acGGA good for finite $r_s$ we define
        \ben
           \epsilon\c\acGGA(r_s,t) = \epsilon\c\LDA(r_s) + H\c\acGGA(r_s,t),
        \een
        where
        \ben
            H\c\acGGA(r_s,t)=\gamma \ln \left(1 + \tilde T(t)^2 f\c(\tilde y)\right ),
           \label{HcacGGA}
        \een
        \ben
           \tilde T = \sqrt{P(t)}T.
           \label{ptacGGA}
        \een
        Enforcing the low density finite-$t$ limit~[\ref{eqars}] now requires
        \ben
            \tilde y = a(r_s) \tilde T(t)^2.
            \label{ypbetilde}
        \een
        That is, we have simply replaced $T$ by $\tilde T(t)$ everywhere in the
        PBE.  We now have an acGGA that meets all the constraints previously
        met by PBE as well as the new condition of asymptotic correctness
        under Lieb-Simon scaling to $Z\to \infty$.

\begin{figure}[htb]
\includegraphics[width=0.85\columnwidth]{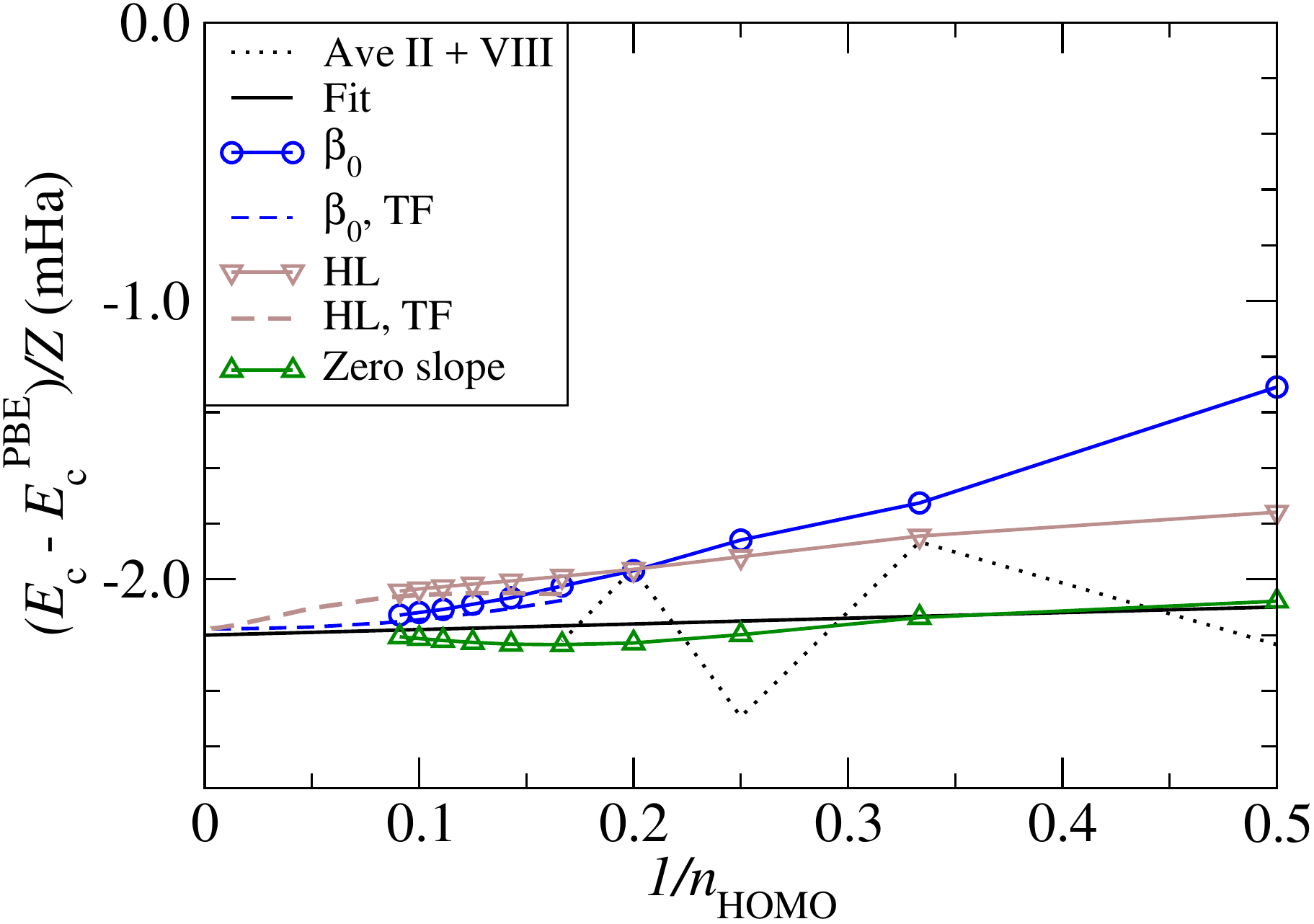}
\caption{
Difference between the correlation energy-per-electron of the acGGA and
PBE averaged over noble gas and alkali earth atoms, plotted versus $1/\nh$,
compared to asymptotic extrapolation from QC data.
Dotted black line shows QC average for $\nh=2$ through $\nh=6$, excepting Ne.
Solid black is the smooth asymptotic curve Eq.~(\ref{eqdelBc}).
Blue circles are acGGA, using the $r_s=0$ value of $\beta$, evaluated
self-consistently through $\nh=11$; blue dashed line, their extension
to $\nh\to\infty$ on the TF density.
Brown triangles and long-dashed line, acGGA-HL, using the
Hu-Langreth $\beta(r_s)$; green
triangles, acGGA+, a modification of HL with $d\beta/dr_s = 0$ at $r_s=0$.
}
\vskip -0.3cm
\label{figextrapolate}
\end{figure}

To see how well the acGGA reproduces the smooth asymptotic
trend defined by Eq.~(\ref{eqdelBc}),
we first plot this trend versus $1/\nh$ in Fig.~\ref{figextrapolate}.
The difference between QC and PBE correlation energies per electron 
averaged over closed shells in each row -- the data to which this trend is fit
-- is also shown to give a sense of the error of the fit.
We compare these to the difference between acGGA and PBE correlation
energies per electron averaged over closed shells in the same way as the
QC data.  These are computed self-consistently up to $\nh = 11$,
and an extrapolation to $\nh \to \infty$ is done by calculating this
averaged energy difference using the Thomas-Fermi density.
These are shown in Fig.~\ref{figextrapolate} as blue circles and blue
dashed line, respectively.  Energies determined using the Thomas-Fermi density 
clearly converge to the extrapolated $B\c$ value in the $\nh \to \infty$
limit, and are very close to the self-consistent ones for large $\nh$.
This provides confirmation that the self-consistent acGGA is in fact
trending to $B\c \sim 37.1$~mHa as designed.

We also note how close the acGGA data is
to a smooth curve after performing our averaging process
-- the effects of shell structure are more than an order
of magnitude smaller than that of the averaged QC data.  This validates our
intuition that the appropriate norm to match a GGA against is not
the atomic data itself, even when restricted to a single column of the
periodic table, but the smooth asymptotic trend derived from that data.

However, while the acGGA correction faithfully follows the
asymptotic trendline at the highest densities, at finite densities
it gradually lifts off the trendline
deviating especially in the ``last" three rows of the ``inverse" periodic
table where it is off by a fraction of a mHa per electron.
Simply fixing $B\c$ removes 90\% of the difference between
PBE and our smooth asymptotic trend for $\nh=6$, but only 60\% for $\nh=2$.

The modest failure of our first try at an acGGA has a relatively
easy explanation and fix.
It is the necessary connection between the modified variable $\tilde T$ used to
generate the high density limit of $H\c\acGGA$ and the
modified variable $\tilde y$ used
in the cutoff function $f\c(\tilde y)$ that determines when PBE crosses over
to its low density, finite-$t$ limit.
At highest $Z$, when $r_s$ is nearly but not exactly zero, the cutoff
function $f\c$ makes very nearly no change to the correlation energy.
This leads
to the flat plateau seen in Fig.~\ref{figextrapolate} for
$E\c\acGGA - E\c\PBE$ as $1/\nh \to 0$.
When $r_s$ gets sufficiently small, the replacement of $y$ in PBE
by $\tilde{y}$ in the acGGA results in a weaker cutoff
because $\tilde T$ has been made smaller than $T$ in order to
reduce $B\c$ from the PBE value.
And thus, on average, PBE correlation will shut off faster than the acGGA,
leading to the rise of the latter relative to the former.

To improve the behavior of the acGGA at finite $r_s$,
a sufficient step is to impose more carefully
the $r_s$ dependence of the GE for correlation,
left unimplemented in PBE.
This correction yields an $r_s$-dependent $\beta$ coefficient to
the gradient expansion [Eq.~(\ref{eqMB})], with $\beta(0)$ equal to the Ma-Brueckner value.   It has been
calculated by two groups,~\cite{HL85, RG86} yielding similar results.
This $r_s$ dependence is rather modest (as shown below) but recent
meta-GGAs~\cite{PRCC09, SRP15} have found it useful
for fine-tuning correlation.
At the level of fine-tuning remaining to adjust the acGGA, it proves to be a
significant effect.

\begin{figure}[htb]
\includegraphics[width=0.85\columnwidth]{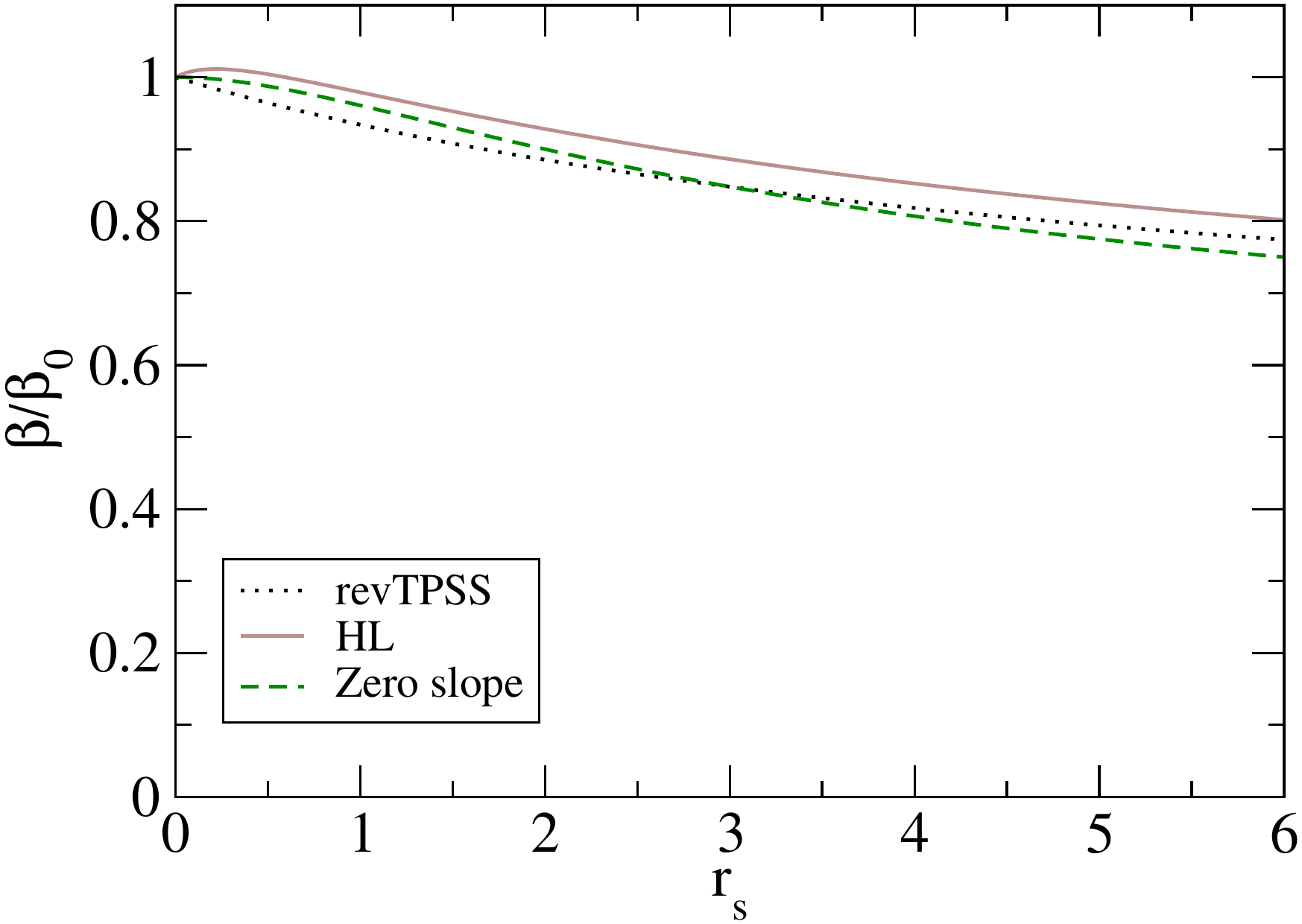}
\caption{Relative variation in the gradient expansion coefficient
$\beta$ as a function of $r_s$.  revTPSS is the model introduced in
Ref.~\onlinecite{PRCC09}.  
The other two are implementations of Eq.~\ref{eqbetars} 
discussed in the text: HL reproduces the model of Ref.~\onlinecite{HL85}
and ``Zero-slope" is designed for close reproduction of Eq.~\ref{eqdelBc}.
}
\vskip -0.3cm
\label{figbetamodels}
\end{figure}

The original Hu-Langreth (HL) form is numerical but we parametrize it roughly
along the lines used in revTPSS~\cite{PRCC09} to obtain
       \ben
         \beta(r_s)=\beta(0)\frac{1 + ar_s(b + cr_s)}{1 + ar_s(1 + dr_s)}.
         \label{eqbetars}
       \een
The coefficients $a\!=\!3.0$, $b\!=\!1.046$ and $c\!=\!0.100$
approximately match the HL form for $r_s<1$.  
The high-$r_s$ limit for $\beta$, however,
is unlikely to be that given by the HL calculation, and instead
we use the limiting condition defined by revTPSS,
setting the ratio $c/d = 1/1.778$.
We also consider a model with zero slope in $\beta(r_s)$ as $r_s\to 0$,
closer in form to that of Ref.~\onlinecite{RG86}, with coefficients
$a\!=\!0.5$, $b\!=\!1$, $c\!=\!0.16667$, $d\!=\!0.29633$.
These models for $\beta(r_s)$ are shown in Fig.~\ref{figbetamodels}, compared
to the one used in revTPSS.  They roughly compare in slope but differ
somewhat in magnitude because of the differing behavior near $r_s\!=\!0$.


The effect of $r_s$ dependence in the GE is to alter the high-density limit
of the acGGA to the form
      $ 
      H\c^{asy}(r_s\to 0, t)=\gamma \ln{\left(1 + \tilde T(r_s,t)^2 \right)},
      $ 
where 
      \ben
           \tilde T(r_s,t) = \sqrt{P(t)}\sqrt{\beta(r_s)/\gamma}\,t 
           \label{eqtildeTrs}
      \een
is the same as $\tilde T(t)$ [Eqs.~(\ref{ptacGGA}) and~(\ref{eqT})] but now using an $r_s$-dependent expression for $\beta$.
A similar change to $\tilde y$ adjusts the transition to the low density form.  The key here is that this generalizes $B\c$ into a weak function of $r_s$.
For high $Z$, the slope of $B\c(r_s)$ is
linearly proportional to that of $\beta(r_s)$ and this offers a way to
tailor the acGGA's functional dependence on $r_s$.


As
$\beta(r_s)$ generally tends to decrease, the outcome for either the HL
or zero-slope model is to lower the effective $B\c(r_s)$ of the acGGA relative 
to PBE.  This pleasingly cancels the trend away from our target asymptotic 
line, so we end up more closely matching QC correlation energies
in the first few rows of the periodic table, as shown in 
Fig.~\ref{figextrapolate}.  However, in the HL gradient expansion,
the slope in $\beta(r_s)$ at $r_s=0$ is \textit{positive},
increasing $\beta(r_s)$ at high density
and lifting the modified acGGA off the asymptotic line
used to measure $B\c$.  This lift makes it impossible to match the
asymptotic line without readjusting $B\c$ by at least a few tenths
of a mHa.  In contrast, the model with zero slope at $r_s=0$ almost 
perfectly matches the asymptotic line.
We thus take the zero-slope model for $\beta(r_s)$ applied in 
Eqs.~(\ref{eqtildeTrs}) and~(\ref{HcacGGA}) as a modified acGGA, denoted
acGGA+.

\ssec{Asymptotically correct exchange}

In order to minimize the overall error in XC,
we apply asymptotic methodology to exchange as well.
There is a fundamental difference between $\zeta$-scaling of exchange and
correlation.  The parameter $s^2$ that determines the gradient correction
for exchange scales to zero as $\zeta\to\infty$, while $t^2$ is 
invariant under $\zeta$-scaling and even at $\zeta\to\infty$ spans a
wide range of values seen in Fig.~\ref{figBct}.
Thus the asymptotic limit of exchange may be used to generate 
appropriate coefficients for a gradient expansion, but does not inform the
entire character of a GGA as we have been able to do for correlation.
Thus one finds the lowest order coefficient for exchange to be $\mu$=0.2603,
in contrast to the formal gradient expansion result of $10/81$.
and recent asymptotic analysis suggests a fourth order 
correction of $-0.125 s^4$.~\cite{CTSCF16}
Notably, any exchange functional that predicts accurate energies for atoms
uses a value of $\mu$ close to that predicted by asymptotic analysis, with
small variations to capture higher order
expansion terms for finite-$Z$ atoms. Conversely, asymptotic analysis
is irrelevant to the large $s$ limit of exchange GGAs, and exchange
functionals with very different behavior in this limit can have 
desirable thermochemical properties.~\cite{PCGTV16,CTSCF16}

Most exchange
functionals, including the commonly used B88~\cite{B88} and PBE, are 
already reasonably asymptotically accurate for exchange.~\cite{EB09}  
For simplicity, we limit our study to these two forms.
Table 1 of Ref. \onlinecite{EB09} shows a small underestimate in the 
coefficient from PBE, but we can correct for this by increasing 
$\mu$ in the formula for $E\x\PBE$ by 13\%, to 0.249.  
We label this acPBEx, denoting modified PBE exchange.
Either acPBEx or B88 make an attractive candidate to pair with acGGA 
correlation, so
we test both forms below.  We will take B88 exchange plus acPBE correlation to 
be the normative acGGA, B88 with acGGA+ correlation as acGGA+, 
and label acPBEx combined with acGGAc as P-acGGA.  

\sec{Measurements and Tests}
\label{Results}
In this section, we take the final acGGA formulas and show their
errors on the neutral atoms (for which they've been designed to be
increasingly accurate with increasing atomic number).  But we also test
acGGA on atomization energies, including attempts to construct hybrids
from acGGA.

    \ssec{Atoms}
    \label{Atoms}
We first explore the behavior of the acGGA and acGGA+ 
across the entire periodic
table.  Complete quantum chemistry data is available for the first
four rows $p = 1$ to 4 of the periodic table but only for closed shells
for $Z>54$.
To augment the available test set for $p=5$ and 6, we replace
the QC data for closed shell atoms with asymptotically corrected
RPA (acRPA) data~\cite{BCGP16} that very nearly duplicates it, and fill
in acRPA data for the open-shell atoms in these rows.
The errors in acRPA data are shown in Table I, and are much smaller than
the difference between acRPA and any functional tested.
For reference exchange energies, we take EXX calculations
using the OPMKS code.~\cite{ED99}

The left side of Table \ref{XCmae} lists errors averaged over row
of the periodic table for
atomic correlation energies with respect to this reference set.
LDA overestimates by about 1 eV per electron, consistent with its error
for $B\c$.  PBE reduces this error by about a factor of 10, consistent with
its almost exact value for $B\c$. But, by being exact for $B\c$,
acGGA reduces this error by a further
factor of 2.
The empirical LYP does best for $Z < 10$, vital to organic chemistry,
but is substantially worse past period 3.  We see the density
dependence in acGGA+ yields no overall improvement relative to acGGA,
but does do better for the second row.

\begin{table*}[t]
\begin{tabular}{|c|c|c|c|c|c|c||c|c|c|c|}
\hline
 & \multicolumn{6}{c||}{$E\c$} & \multicolumn{4}{c|}{$E\xc$}\\
\hline
p & acRPA & LDA & LYP   & PBE   & acGGA & acGGA+& PBE & P-acGGA & acGGA & acGGA+ \\
\hline \hline
1 & N/A & 0.765 & 0.011 & 0.084 & 0.094 & 0.112 & 0.216 & 0.032 & 0.039 & 0.037 \\
2 & N/A & 0.924 & 0.024 & 0.067 & 0.038 & 0.032 & 0.304 & 0.018 & 0.080 & 0.070 \\
3 & N/A & 1.032 & 0.047 & 0.045 & 0.014 & 0.018 & 0.297 & 0.104 & 0.023 & 0.013 \\
4 & N/A & 1.002 & 0.082 & 0.113 & 0.061 & 0.055 & 0.355 & 0.114 & 0.016 & 0.014 \\
5 & 0.003 & 1.082 & 0.107 & 0.055 & 0.010 & 0.010 & 0.433 & 0.083 & 0.010 & 0.013 \\
6 & 0.015 & 1.034 & 0.271 & 0.120 & 0.067 & 0.063 & 0.472 & 0.082 & 0.041 & 0.045 \\
\hline \hline
All & N/A & 1.020 & 0.146 & 0.092 & 0.047 & 0.044 & 0.401 & 0.084 & 0.031 & 0.031 \\
\hline
\end{tabular}
\caption{Mean absolute error (eV) of energy components per electron,
taken with respect to our
reference data set, and averaged over each period (p) of the periodic table.
(The reference data set is given in Ref.~\onlinecite{BCGP16}, and consists of QC data for $Z\leq 54$ and asymptotically corrected RPA (acRPA in Ref.~\onlinecite{BCGP16}) for $p=5$ and 6.)  
acRPA is RPA
adjusted to match the asymptotic limit of quantum chemistry data, and used to
fill in $Z$ values in that data for $p=5$ and 6.
}
\label{XCmae}
\vskip -0.25cm
\end{table*}

For XC together, acGGA correlation with acPBEx (P-acGGA) is about 4 times more accurate
for atoms than PBE is.  However, B88 is so accurate throughout the table
as well as asymptotically, that when combined with acGGA correlation (acGGA), its error 
is three times smaller again.  Finally the addition of density-dependence to the 
correlation energy gradient in acGGA+ (B88 exchange and acGGA+ correlation) improves
cancellation of error (relative to acGGA) up to the fourth row, and smooths
out the fluctuations between even and odd rows.

\begin{figure}[htb]
\includegraphics[width=0.85\columnwidth]{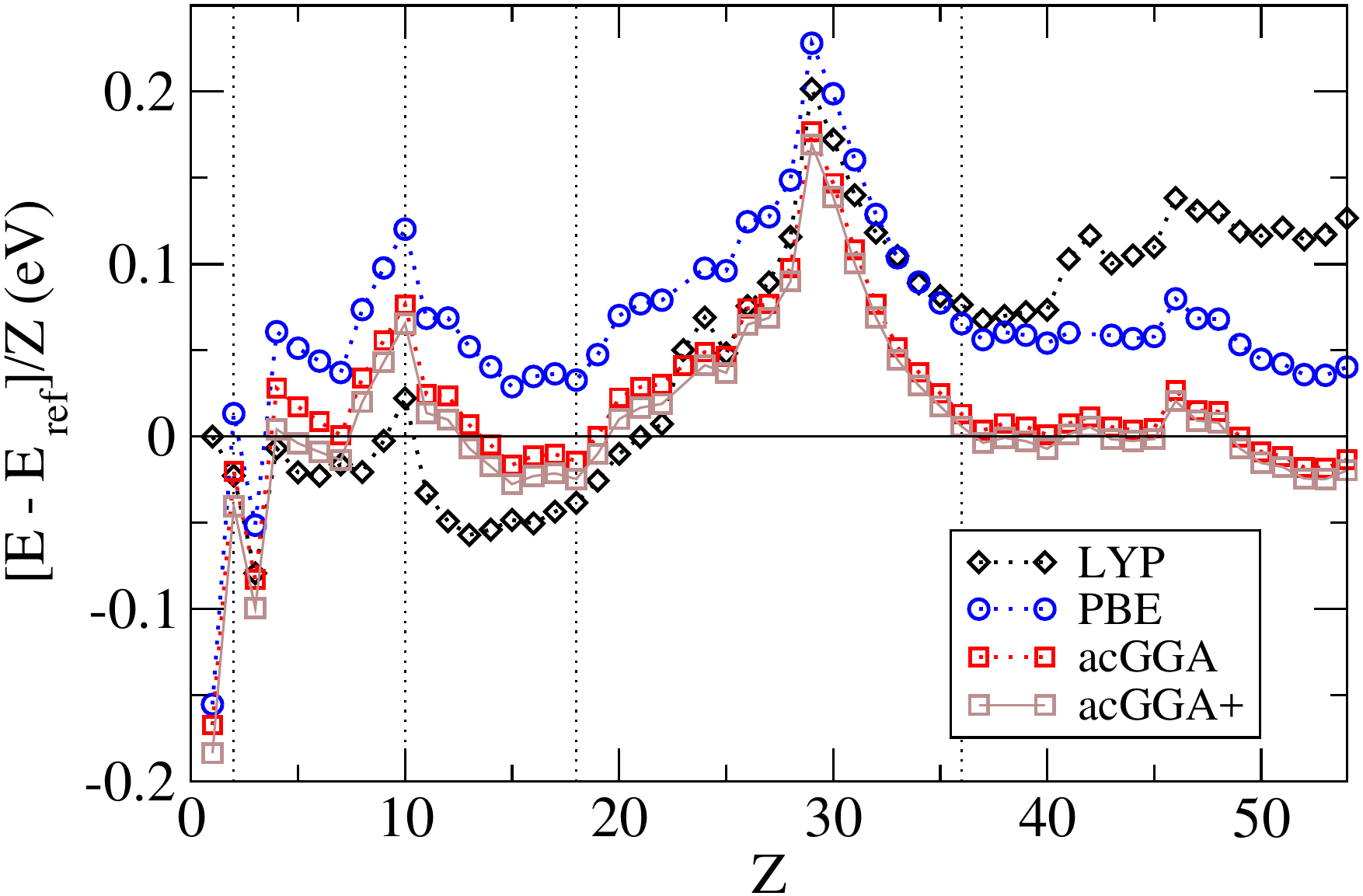}
\caption{Errors in correlation energy per electron as a function
of atomic number.  Closed shells
indicated by vertical dotted lines.
}
\vskip -0.3cm
\label{figCtents}
\end{figure}

We show the difference between density functional and QC correlation energies 
per electron
for atoms with $Z\leq 54$ in Fig.~\ref{figCtents}. 
PBEc is, for much of the periodic table, roughly a constant shift off from
QC reference data except for underperforming regions at the end of the second
row and the middle of the fourth.  The asymptotic correction of the
PBE, acGGA, produces a nearly constant shift with respect to PBE for all $Z$,
indicating that it has a nearly exact representation of the overall general
trend of correlation energies with $Z$ but is no more sensitive
to the details of shell structure than is PBEc.
The $\beta(r_s)$ correction included in acGGA+ is a
small perturbation upon these results, but
as one might expect, is a noticeable improvement in the second row.
The LYP correlation functional has an error that in addition to the uncontrolled growth with
$Z$ noted earlier, has rather large fluctuations even for lower rows of
the periodic table.

\begin{figure}[htb]
\includegraphics[width=0.85\columnwidth]{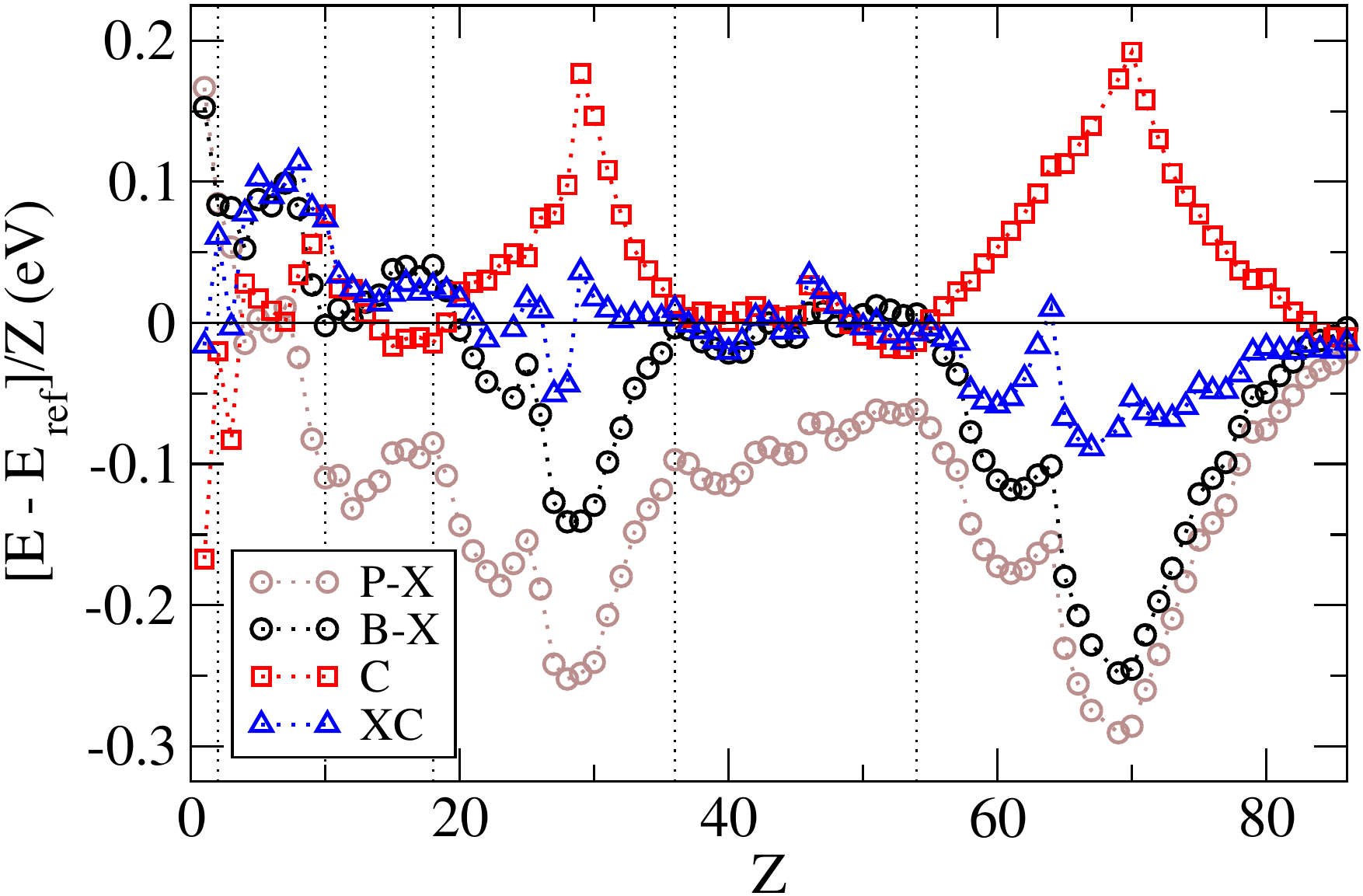}
\caption{Errors in XC components per electron as a function
of $Z$ for acGGAc (C), acPBEx (P-X), B88 exchange (B-X),
and B88 exchange with acGGAc (XC).
PBE errors are
significantly larger (see Table \ref{XCmae}) and do not often cancel.}
\vskip -0.3cm
\label{figXCtents}
\end{figure}

In Fig.~\ref{figXCtents} we show errors in energy per electron for
acGGA correlation,  acPBEx, B88 exchange,
and the combination of B88 with acGGA correlation.
We note an eerie match of B88 exchange with acGGA correlation --
both are exceptionally accurate for odd rows and exhibit a strong
anticorrelation of error in even rows.
The X and C errors are
like mirror images, so that they largely cancel one another, just as in LDA,
making XC much more accurate than X.
The worst actors ($Z=10,29,30,70$)
are {\em the same} for both X and C.
The cancellation of X and C errors
likely is attributable to a cancellation between X and C holes, as the latter is
affected by the screening characteristics of the former.
(A classic example is the long range tail in the 
exchange hole in a uniform gas inducing a long
range tail in correlation hole which cancels the effect.  The effect is
to decrease the magnitude of the LDA exchange energy and increase that of LDA 
correlation relative to the exact values for any non-metal or finite system.)

To begin to understand why the bad actors are who they are,
note that asymptotic expansions fare worst when only the lowest level
of a quantum system is occupied,~\cite{CLEB10}
This happens here for each angular momentum, $l$.
Consider $\n(\br)$ as a sum of contributions with different angular
shapes, $n_l(\br)$.
Whenever a given $l$ value is first occupied, our errors should be largest.
In the first octet, the lowest $p$ orbitals
are occupied (first singly, then doubly) across
the row, leading to the
largest error when full (Ne).  The problem slowly goes away
in the second octet, as each channel gains a 3$p$ occupant,
but recurs when first filling the $d$ orbitals, being
worst for closed 3$d$-shell atoms ($Z=29$) and Zn ($Z=30$), and again for
the $f$-orbitals at Yb ($Z=70$).
This is only a partial explanation of the phenomenon
because the error recedes not with the first introduction of the second
shell with the same angular momentum, but when the first is sucked
into the core with the introduction of additional valence shells.
Possibly, the spatial isolation of such shells
when in the valence amplifies their deviation from asymptotic behavior,
but a more detailed explanation requires further research.

Finally, we note that  acPBEx has qualitatively the same
behavior as the B88 form, but never does quite as well as it in matching
EXX energies.  It noticeably approaches
B88 as $Z$ gets larger, naturally, because the two forms
have the same $B\x$ and must converge as $Z\to \infty$.

\ssec{Molecules}
To test the effect of our asymptotic correction to the GGA for
practical applications, we
look at atomization energies of molecules -- specifically
those of the HEAT~\cite{TSCK04} and G2-1~\cite{CRRP97} data sets.
Although we take an asymptotic analysis in the $Z\to\infty$
limit, a good asymptotic expansion is useful for any system with $Z^{-1}<1$.
An improved asymptotic analysis should therefore provide a noticeable
benefit for the thermochemistry of organic
molecules -- for which the typical values of $Z^{-1}$
of many of the constituent
elements are less than 0.2.
We evaluate the approximate functionals on PBE orbitals as these systems
are normal and the results should change little under self-consistency.
All DFT calculations have been performed using a modified version of
Turbomole 6.6.~\cite{TURBOMOLE14}  Atom-centered Gaussian basis sets of valence
quadruple-zeta plus polarization quality (def2-QZVP) are used for all
atoms.~\cite{Weigend03} A fine density grid of quality 6 was employed for
numerical integration.~\cite{Treutler95} The accuracy of different XC functionals was
assessed for atomization energies using HEAT and G2-1 test sets. The
results are compared with high-level coupled-cluster (CCSDTQ)~\cite{TSCK04}
and CCSD(T)~\cite{FP99} calculations, and tabulated in the supplementary material for this article.

In Table~\ref{Molecules} we show mean absolute errors, median errors,
and maximum spread or difference between the most positive and most negative 
errors, across the HEAT and G2-1 test sets.
The median error shows that PBE and BLYP~\cite{B88,LYP88} have a 
systematic tendency to
overbind, although they are a great improvement on the LDA which has a
median overbinding of 38~kcal/mol.
Both flavors of the acGGA reduce the overbinding of PBE.
B88~\cite{B88} exchange plus acGGA correlation (acGGA) has the best median
value overall, cutting the overbinding error in PBE in half for the HEAT set,
and even more dramatically for the G2-1.
It has a somewhat larger maximum spread of errors compared to the LYP --
it is less successful at improving the precision of
PBE calculations than in correcting its median error.  This leads to
a somewhat larger MAE than that of BLYP.
As with atoms, the inclusion of an $r_s$-dependent $\beta$,
or acGGA+, improves only slightly upon the acGGA, indicating that the latter
is already nearly optimal.
The P-acGGA, using 
asymptotically correct acPBEx, has much less, but non-negligible, effect.

\begin{figure}[htb]
\includegraphics[width=0.85\columnwidth]{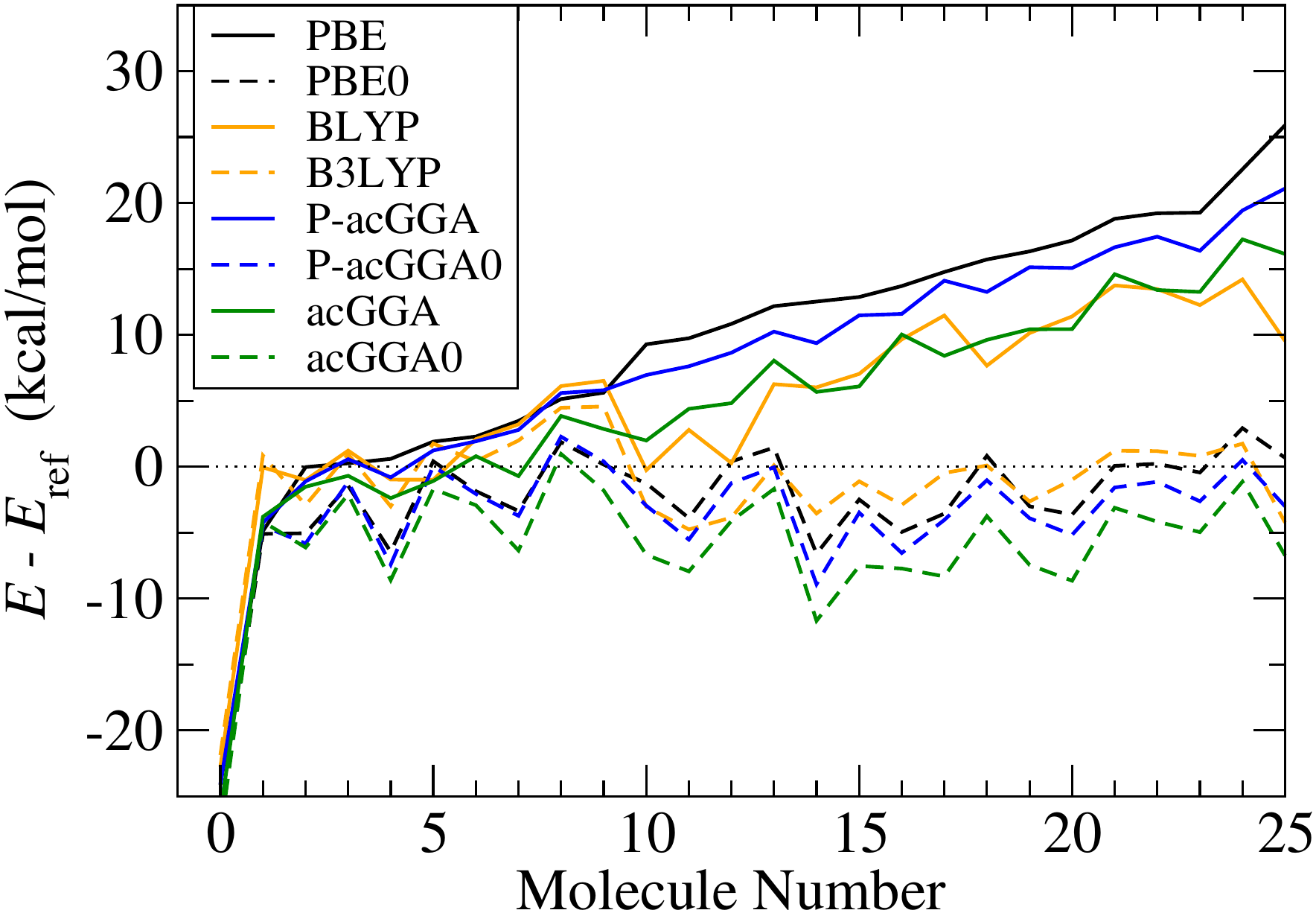}
\caption{Atomization energy errors across the HEAT data set, sorted via
PBE errors.  P-acGGA is acGGA correlation with asymptotically corrected
PBE exchange, acGGA with B88 exchange.  Hybrids P-acGGA0 and acGGA0
evaluated with 25\% mixing of HF with semilocal exchange.
}
\vskip -0.3cm
\label{figHEAT}
\end{figure}

\begin{table}[htb]
\begin{tabular}{|l|r|r|r|r|r|r|}
\hline
\hline
      &  \multicolumn{3}{c|}{HEAT}  &  \multicolumn{3}{c|}{G2-1}  \\
\hline
Model & ~MAE~ & ~ME~ & MS & ~MAE~ & ~ME~ & MS \\
\hline
BLYP & 6.97 & 6.19 & 36.96 & 5.27 & 1.66 & 30.75\\
PBE & 11.51 & 11.51 & 50.02 & 8.52 & 5.33 & 44.29\\
P-acGGA & 10.10 & 9.01 & 45.16 & 7.25 & 3.96 & 40.72\\
acGGA & 7.64 & 5.25 & 43.52 & 5.88 & 1.41 & 38.95\\
acGGA+ & 7.53 & 5.26 & 42.89 & 5.71 & 1.60 & 37.29\\
\hline
B3LYP & 2.90 & -0.19 & 26.44 & 2.78 & -0.69 & 21.35\\
PBE0 & 3.40 & -1.64 & 29.22 & 3.45 & -1.42 & 18.17\\
P-acGGA0 & 4.05 & -2.98 & 28.40 & 3.79 & -2.43 & 21.42\\
acGGA0 & 6.08 & -5.54 & 28.76 & 5.46 & -5.15 & 24.40\\
P-acGGAopt & 3.12 & -0.45 & 30.01 & 3.48 & -1.73 & 19.85\\
acGGAopt & 3.49 & -0.60 & 34.08 & 3.79 & -2.18 & 24.91\\
acGGA+opt & 3.37 & -0.65 & 33.45 & 3.59 & -1.94 & 23.61\\
\hline
\hline
\end{tabular}
\caption{\label{Molecules}
Mean absolute error (MAE), median error (ME) and maximum spread (MS) of
atomization energies of GGAs and hybrid functionals
across 26 molecules of the HEAT data set~\cite{TSCK04} and 55 molecules of the
G2-1 set,~\cite{CRRP97} in kcal/mol.  
Atoms have been excluded in both cases.
}
\end{table}
In Fig.~\ref{figHEAT},
we show errors in atomization energies of the HEAT test set 
for several GGA and hybrid functionals, sorted by increasing size of PBE errors.
The PBE errors strictly separate into three groups: molecules 0 -- 9, each
of which have a single non-hydrogen atom, 10 -- 24, which have two, and
molecule 25, carbon dioxide, which has three non-hydrogen atoms and the
largest error.
They group only crudely along the number of electrons in the molecule or
other measures.
The PBE is already very good for the first set, and the acGGA only slightly
improves upon it.  There is a definite improvement for P-acGGA when one
moves to the two non-hydrogen atom set, and most improvement for CO$_2$.
The same pattern is followed by the acGGA and BLYP which closely match
each other on a per-atom basis. 
For the G2-1 data set, the same trend occurs -- P-acGGA is only a
minimal change from PBE for molecules with only one
non-hydrogen but a noticeable improvement of about 10~kcal/mol for two
such atoms and even more for three.
Conversely, the small improvement provided by the acGGA+ for low $r_s$
shows up only for the H-rich molecules where acGGA is least effective, 
leading to the lower maximum spread shown in Table~\ref{Molecules}.
This pattern is consistent with our
hypothesis that the asymptotic correction of a functional is relevant
for any $Z>1$, as it
has the most noticeable effect for molecules in which
second row atoms and not hydrogen are the dominant players.

\begin{figure}[htb]
\includegraphics[width=0.85\columnwidth]{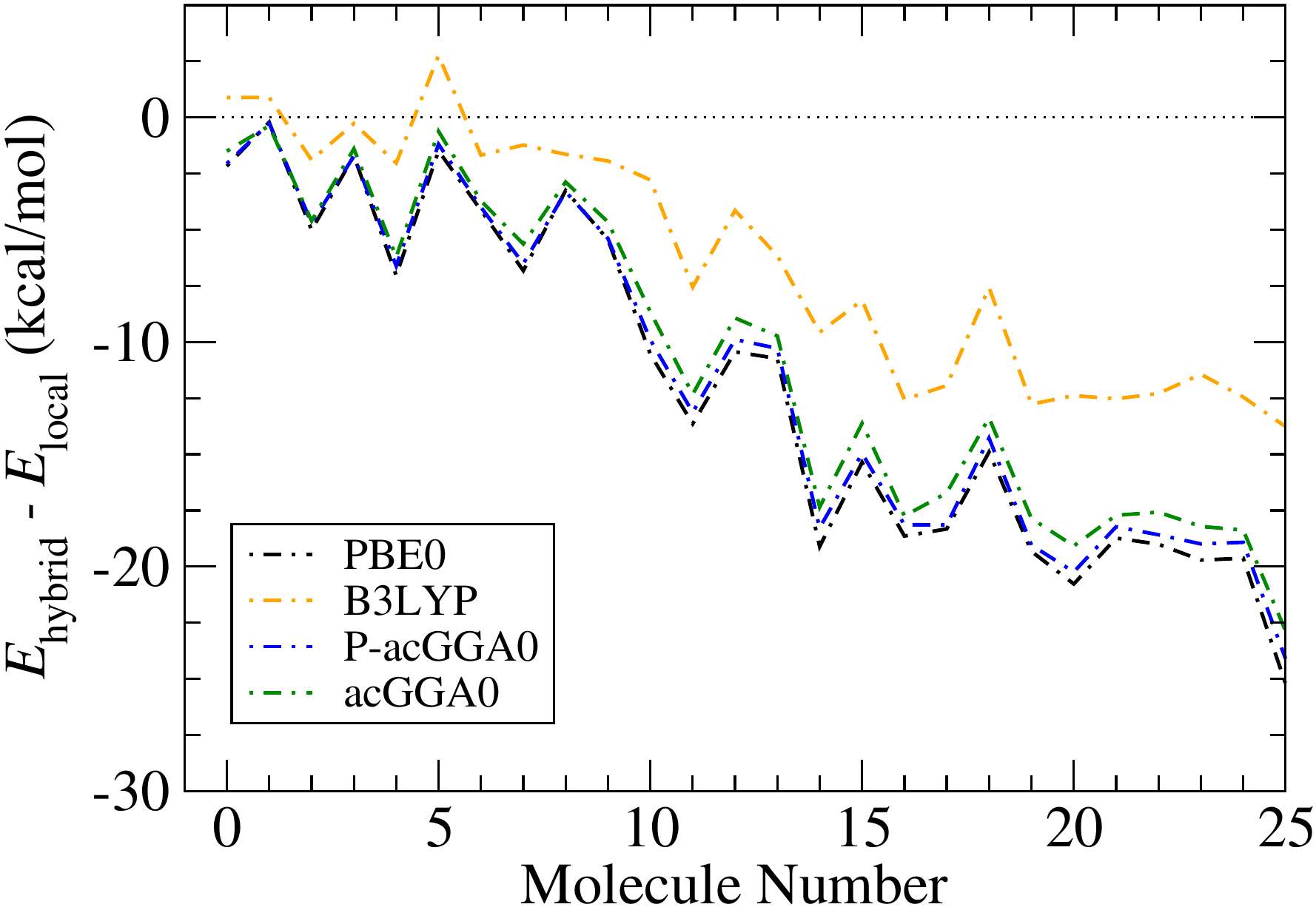}
\caption{
Difference across the HEAT data set between the atomization energy evaluated
with various hybrid functionals of Fig.~\ref{figHEAT} and the related semilocal
functionals.
}
\vskip -0.3cm
\label{figdelHEAT}
\end{figure}

With the hybrids, the story is different.  We calculate
parameter-free ``DFT0" hybrids, made by 
combining strictly 25\% HF exchange with DFT exchange,~\cite{PEB96} 
and compare to common 
hybrids PBE0~\cite{PEB96, BEP97} and B3LYP.~\cite{B93,SDCF94}
The empirically fit B3LYP is the best of this class, and no acGGA
25\% hybrid improves upon PBE0.
Fig.~\ref{figdelHEAT} shows that 
the asymptotic corrections we have made to the acGGA, either exchange or
correlation, have little effect on the size of hybrid correction
as long as a fixed 25\% mixing is taken.
The 25\% hybrid correction in PBE0 is nearly optimal,
with a small amount of underbinding in the median for
both HEAT and G2-1.  The underbinding correction of the acGGA that makes it the
best overall GGA also makes it the worst 25\% hybrid, as the median errors for
each version are shifted down by almost exactly the same amount
upon hybridization.  As a result, MAE's are reduced by much less than one
might hope for with 25\% mixing.
However, one
can reproduce or slightly improve
PBE0 MAE for empirical acGGA hybrids, by mixing
a smaller fraction of HF exchange (20\% HF exchange with
acPBEx exchange or 14\% with B88). 
This is similar to hybrids of meta-GGAs,
such as the TPSSh, the hybrid of the TPSS meta-GGA and exact exchange, which is
optimized at 10\% mixing.~\cite{SSTP03}  The results (P-acGGAopt and acGGAopt)
are then close to those of PBE0.
Hybrids formed from the acGGA+ are optimized with the same amount of mixing
as those formed from the acGGA and are again only a small improvement on 
the latter.
In all, acGGA ought to be a better starting point than PBE, requiring smaller
fractions of HF exchange to produce accuracies similar to PBE0.

\sec{Implications and Conclusions}
\label{Implications}

In this section, we discuss the implications of
our effort to construct an asymptotically correct density
functional.

\ssec{Relevance of the $Z\to \infty$ limit}
The main implication of our work is the demonstration for the correlation
energy, of the importance of $\zeta$-scaling; in particular,
the importance of the $Z\!\to\!\infty$ limit for all neutral atoms with $Z\!>\!1$.
We see that the asymptotic expansion for correlation derived in our previous
work, implemented into an asymptotic functional for the $r_s\!=\!0$ limit
of the correlation energy, by itself accounts for 80\% or more of the
beyond-LDA correlation for closed shell atoms down to Be.
Integrating this limit with other constraints,
most notably the low density, finite $t$ limit of the
PBE, generates a functional that is highly accurate for nearly all atoms.
It seems to be the interplay of constraints, that necessarily
become ``entangled" with each other that does this;
the form of the asymptotic limit of the GGA imposes specific
conditions on the nature of the correlation cutoff function $f\c$,
and thus propogates information on the $r_s=0$ limit to the functional
at all $r_s$.  The improvements are thus not limited to heavy
atoms, but are significant even in the second row of the periodic table.
Most notably, the asymptotic limit has an effect on the bonding characteristics
of small-$Z$ molecules -- atomization energies are noticeably improved over the
PBE whenever there are bonds between two or more $Z > 1$ atoms.

One question that our work does not quite resolve is
why does PBE correlation parallel the beyond-$B\c$, low-$Z$
behavior of QC so well?
PBE correlation seems uncannily successful --
satisfying constraints in limiting cases need not guarantee
the level of accuracy of PBE (and therefore of the acGGA) for
intermediate situations.  
As an illustration of this point, note the
difference in performance of the Pad\'e and B88 forms for acGGA exchange.
Both meet the same constraints in the limit of large $Z$ and are reasonable
parametrizations of the RSC exchange energy for moderate levels of
inhomogeneity, but the B88 is clearly superior in
in recovering atomic exchange-correlation energies and molecular atomization
energies.  It could be argued that in each case (PBE correlation and B88
exchange) one uses the two best possible constraints for
$Z\sim 1$ and $Z\to \infty$ systems, which then is sufficient to nail
down the energies of most atoms.
The Pad\'e form for PBE exchange was designed, as much as possible,
for universal applicability (here, ensuring the global Lieb-Oxford
bound for any system) and not for optimal behavior for a specific class of 
systems.

\ssec{Functional development}

We briefly consider the implications of our work for future functional 
development.

Despite our optimization of the basic form of the GGA for atomic energies,
we expect that there is nearly as much room in functional space for
tuning GGA correlation as there has been for exchange.
The ``internal enhancement factor" $f\c$
used to modify the high density form of PBE is
as open to variation as the enhancement factor $F\x$
is for exchange.
The two functions incorporate an intriguingly similar scaling form; and
though not completely invariant under uniform scaling, the scaled variable $y$
used in $f\c$ is close to $s$, the invariant argument used in exchange.
Each thus describes a transition between small-$s$ and large-$s$ limits.
At the same time, the large-$r_s$ form for $\beta(r_s)$ is open to
improvement, as it is largely unknown.
$f\c$ and $\beta$ may be manipulated together to come up
with an infinite variety of forms that preserve asymptotic correctness
and the correlation energies of atoms at finite $Z$, while meeting
other conditions, perhaps on the potential.  

Our work naturally also has consequences for higher rungs of functional
development, as we have already discussed in regards to hybrids.
The standard next step beyond the GGA in functional development is
the meta-GGA which adds information obtained from the local Kohn-Sham 
kinetic energy, $\tau\KS$, in addition to the local density and gradient,
A common approach to meta-GGA development
is to parametrize corrections to the GGA in terms of an electron
localization measure,~\cite{SHXB13,SRP15}
    \ben
        \alpha = \frac{\tau\KS - \tau\W}{\tau\TF}
    \een
where $\tau\KS \!=\! (1/2)\sum_i^{Nocc} \left | \nabla \phi_i \right |^2$,
and is calculated with occupied KS orbitals $\phi_i$,
$\tau\W \!=\! |\nabla \n|^2/8\n$ is the von Weizsacker kinetic energy functional
and $\tau\TF$, the TF energy density.
This measure is closely related to the electron localization
factor (ELF),~\cite{BE90,SS94b} and like the ELF, distinguishes between
three limit cases.
The limit $\alpha=0$ indicates single orbital occupation, typified by
covalent single bonds, $\alpha=1$ the highly-degenerate
electron gas, and thus metallic bonds, and $\alpha\to\infty$,
regions of asymptotically low electron densities such as ionic bonds.
Typically, a constraint-based meta-GGA tries to handle each case with a
different exchange-correlation functional.~\cite{SRP15}

Within the context of meta-GGAs, our work is immediately relevant to the
high-density, high-degeneracy limit, or $\alpha\!=\!1$ and $r_s\!=\!0$,
the limit of $\zeta\!\to\!\infty$ in Lieb-Simon scaling.
In order to reproduce the correct $B\c$ coefficient in the asymptotic
expansion for correlation, the correlation functional
should reduce to something like Eq.~(\ref{PBEnewasy}) in this limit.
In addition, the beyond-$B\c$ asymptotic
trend of Eq.~(\ref{eqdelBc}) involves a transition from $\alpha\!=\!1$ for
$Z\to\infty$ to $\alpha\!=\!0$ as one reaches the He spin singlet.
Matching the correlation energies of closed shell atoms for finite $Z$
could be an appropriate norm for determining this transition, and this
is used in the construction of the
recent constraint-based SCAN meta-GGA functional.
SCAN's reliance upon non-asymptotically corrected PBE correlation implies an
inaccurate
value for $B\c$, but on a scale that is likely irrelevant to the resulting
approximation. 

The approach taken in our work may also be useful beyond its 
immediate scope --  to analyze directly the standard ingredients of meta-GGAs.
As we have noted earlier, Lieb-Simon scaling analysis has produced an 
estimate of the fourth-order gradient correction of exchange in atoms~\cite{CTSCF16}; in addition it has been used to deduce the 
large-$Z$ limit of $\alpha$ for atoms, and 
to show that it has relevance for physical values of $Z$.~\cite{CR17}

\ssec{Limits of the asymptotic correction}
Perhaps the most interesting physical issue raised by our work 
is the separation of the periodic table into sections that are close to
the asymptotic limit and others that are not.
We note that this misfit is maximum for \textit{closed-shell} atoms and not,
as one
might expect, for open-shell systems.  The apparently relevant argument as
one fills the 3d shell or 4f shell is the filling fraction of the shell,
not other details such as non-spherical potentials.
Open-shell structural effects are in comparison only
responsible for low level ``noise" in the overall trend.
To account for the majority of the remaining exchange and
correlation error in atoms, a next-order correction to the acGGA need,
quite contrary to our initial expectations, only look at failure
modes for spherical, unpolarized systems.

Significantly, the little data we have to characterize this trend
shows no evidence that the problem would eventually go away for
very large $Z$.  Our functional is close to ideal for exactly one-half
of the periodic table (odd rows) despite no consideration of any open-shell
system, but still substantially in error for the other half (even).
There is a possibility that there is a dependence of the
asymptotic coefficients $B\x$ and $B\c$ with filling fraction for these bad
rows, not obtainable from an extrapolation that considered atoms only from
near the boundaries of each row.  Such behavior is not unexpected in
asymptotic analysis, appearing for example, in the expansion of
total energy of the Bohr (noninteracting) atom.~\cite{BCGP16}
Careful asymptotic analysis of this trend for exchange, to which correlation
seems to be a response, and careful consideration of functionals that could
model this effect would both be very welcome.

\ssec{Conclusion}
\label{Conclusion}

The central result of this paper is the construction of a density
functional for correlation that satisfies the leading-order
correction to LDA correlation in the 
large-$Z$ limit of neutral atoms  
as determined from the best QC data available.  The functional is
implemented as a modification to the PBE generalized gradient
approximation, the simplest possible level of density
functional at which this asymptotic correction can be obtained.
The importance of this limit for all electronic structure
is shown by its impact on the correlation energies for the
entire periodic table and atomization energies for molecules
of standard thermochemistry test sets.
Together with an asymptotically correct exchange, this functional is
close to the
best parametrization of the energy of atoms across the periodic table that may be constructed at
the GGA level.  Our functional should thus serve as a starting point
and a benchmark for constructing improved meta-GGA and hybrid functionals.

\usec{Supplementary Material}
See supplementary material for the specification of the acGGA functional
and its potential, and for tables of atomic exchange and correlation energies
and atomization energies for the HEAT and G-21 data sets.

\usec{Acknowledgments}
This work was supported by NSF CHE-1464795.
We thank John Perdew and Jianwei Sun for many useful discussions and
Eberhard Engel for use of his atomic DFT code, OPMKS.

\appendix

\section{Divergence of gradient expansion for correlation}
The naive gradient expansion for correlation in the Lieb-Simon asymptotic
limit is
\ben
    E\c^{aGE} = \gamma\ln(r_s) + \eta + \beta(r_s) t^2.
    \label{eq:Z}
\een
where the first two terms give the high-density RPA limit of the LDA,
Eq.~(\ref{epsunif}).
It is not shown in our plots.  It is too large even
for finite atoms, being over 100~mHa for Neon.  Secondly, it diverges
logarithmically for large $Z$.  This is a surprising result,
as $t^2(\br)$ scales as a constant under $\zeta$-scaling. But it may be
explained as follows:

While
$t^2(\br)$ does scale as a constant under $\zeta$-scaling, that constant
tends to infinity at the nucleus because the scaled density does too in
the TF limit.
Take the following convenient expressions~\cite{LCPB09} for the
radial particle density and $t^2$ as $r \rightarrow 0$:
\ben
    4\pi r^2 n\TF(\br) dr \rightarrow Z x^{1/2} dx
\een
and
\ben
    t^2 \rightarrow \frac{a_2^2}{x^{3/2}}
\een
where
\ben
    x = Z^{1/3}r/a
\een
and $a = (1/2)(3\pi/4)^{2/3}$ and $a_2 \sim 0.6124$.
The expression for the GEA contribution to the energy in this limit is
\ben
    E\c\GEA = Z a_2^2 \int_0 x'^{-1} dx' \sim \ln(x) |_0
\een
For a finite $Z$ system the logarithmic divergence is cured by the
transition to the nuclear cusp occurring around $r = a_0/Z$ or
$x = a_0/aZ^{2/3}$.
The density no longer
diverges as $1/x^{3/2}$ but goes to some definite finite value.
If we take the lower limit of the integral over the diverging Thomas-Fermi
density to be $a_0/Z$ this diverging term becomes:
\ben
    E\c\GEA \rightarrow Z a_2^2 \ln{(aZ^{2/3}/a_0)},
\een
or
\ben
   E\c\GEA \rightarrow \frac{2a_2^2}{3} Z\ln{Z}.
\een
Thus,
the GE  produces a finite contribution or order $Z\ln{Z}$
to the asymptotic expansion of the energy of neutral atoms.
This contribution is spurious, as the work of
Kunz and Rueedi~\cite{KR10} already implied that 
the coefficient of this term is exactly given by
LDA correlation.

\section{Derivation of asymptotically corrected $H\c$}
To find an analytic value of $\tilde c$ in \Eqref{eqPt}, we derive
RSC in the large-$Z$ limit.
Appendix C of Ref.~\onlinecite{BPW97c} gives formulas for RSC as $r\s\to 0$.
Both the LDA and GEA correlation holes are given in terms of a short-ranged
contribution (on the scale of 1/$k\F$) and a long-ranged contribution
(on the scale of $1/k\s$).
As $r\s\to 0$, the short-ranged pieces do not contribute.  The long-ranged
radial LDAc hole tends to a constant as $v=k\s u \to 0$, so
the energy integral has a $1/v$ term, the cutoff of which produces the
$\ln r\s$ contribution to the correlation energy in Eq. \ref{epsunif}.
As $t$ becomes large, the cut-off $v\c$ is very small, producing the
logarithmic divergence with $t$.

Although PBE removes the logarithmic divergence, we have seen
it clearly differs from the RSC in the next order.
Define
\ben
    C=\lim_{t\to\infty}\left[H\GGA\c(0,t) - 2 \gamma\, \ln t \right],
\een
to find $C\PBE= \gamma \ln(\beta/\gamma) =0.0237$.   For the real-space
construction, define
\ben
\gamma = \lim_{\epsilon\to 0} \int_\epsilon^\infty dv\, \frac{f_1(v)-4\gamma}{2v},
\een
where $v=k\s u$ and $f_1(v)$ is the dimensionless
radial $\n\c\LDA(u)$ in RPA.  Then
\ben
    C\nGGA = \gamma \left[ 3 - 
             2 \ln \left(3\pi{\sqrt{6 \gamma}}\right)\right] + \gamma,
\een
which is about -0.0044 with the models of Ref.~\onlinecite{BPW97c}.
Then take $\gamma\ln\tilde c\nGGA \!=\! C\PBE-C\nGGA$, to yield
$\tilde c \!=\! 2.4683$.


\begin{thebibliography}{94}
\bibitem [{Der()}]{DerivNote}%
  \BibitemOpen
  \href@noop {} {}\bibinfo {note} {The energy as a function of the smooth
  variable $\zeta$ has complications such as derivative discontinuities at
  integer $Z$. These corrections to $E_{TF}(\zeta)$ become vanishingly small in
  the Lieb-Simon limit. Corrections to $E_{TF}$ at the GGA level yield
  continuous curves that are more accurate at integer $Z$, without addressing
  these discontinuities.}\BibitemShut {Stop}%
\bibitem [{BcN()}]{BcNote}%
  \BibitemOpen
  \href@noop {} {}\bibinfo {note} {This disagrees slightly with our previous
  result,~\cite{BCGP16} reflecting a small error in weighting points in the
  latter.}\BibitemShut {Stop}%
\bibitem [{Cac()}]{CacNote}%
  \BibitemOpen
  \href@noop {} {}\bibinfo {note} {We choose to alter $\tilde c\ac$ rather than
  $\eta$ because the GE hole in the $r_s=0$ limit is well-known\cite{LM83},
  leaving the crude cutoff method employed as the major source of error in the
  RSC. Large-$t$ features of the RSC GGA are thus those most likely to be in
  error.}\BibitemShut {Stop}%
\end{thebibliography}
%

\end{document}